\documentstyle[aps,prc,eqsecnum,epsf,exscale,twocolumn]{revtex}

\draft

\newcommand{\ch}{{\rm ch\,}}
\newcommand{\sh}{{\rm sh\,}}
\newcommand{\arth}{{\rm arth\,}}

\begin{document}
\wideabs{
\title{Coalescence and flow in ultra-relativistic heavy ion collisions} 
\author{R\"udiger Scheibl and Ulrich Heinz}
\address{Institut f\"ur Theoretische Physik, Universit\"at Regensburg, 
         D-93040 Regensburg, Germany}
\date{March 16, 1999 --- 
      published in Physical Review C 59 (1999), 1585-1602}
\maketitle

\begin{abstract}

Using a density matrix approach to describe the process of
coalescence, we calculate the coalescence probabilities and 
invariant momentum spectra for deuterons and antideuterons. 
We evaluate our expressions with a hydrodynamically motivated
parametrization for the source at freeze-out which implements rapid
collective expansion of the collision zone formed in heavy ion
collisions. We find that the coalescence process is governed by the
same {\em lengths of homogeneity} which can be extracted from HBT
interferometry. They appear in the absolute cluster yield via an {\it
  effective volume} factor as well as in a quantum mechanical
correction factor which accounts for the internal structure of the
deuteron cluster. Our analysis provides a new interpretation for the
parameters in the popular phenomenological coalescence model and for
the effective overlap volume in Hagedorn's model for cluster
production in $pp$ collisions. Using source parameters extracted from
a recent HBT analysis of two-pion correlations, we successfully describe
deuteron and antideuteron production data from Pb+Pb collisions at the
CERN SPS as measured by the NA44 and NA52 collaborations. We also
confirm the recent finding by Polleri {\em et al.} that the
different measured slopes of nucleon and deuteron transverse mass
spectra require a transverse density profile of the source which is
closer to a box than to a Gaussian shape. 

\end{abstract}



 } 

\section{Introduction}

Ultra-relativistic heavy ion collisions are used to study in the
laboratory nuclear matter at extreme energy density and
temperature. For such conditions lattice QCD predicts the transition
from a hadron gas to a quark-gluon plasma. The only observables of the
hot, central reaction zone (fireball) created in the heavy ion
collision are the energies and the momenta of the produced
particles. Most of these are hadrons which, due to their strong
interactions, decouple only late from the collision fireball and
therefore carry direct information only about this so-called
``freeze-out stage''. Direct signals from the earlier (presumably much
hotter and denser) stages of the collision are carried by
electromagnetic (photons, dileptons) and ``hard'' probes (jets,
$J/\psi$) which are, however, more difficult to study due to their much  
smaller cross sections. Developing a clear and unique picture of the
interesting early stages of ultra-relativistic heavy ion collisions
will thus require the combination of such direct signatures with an
extrapolation backwards in time of the information extracted from the
spectra of the bulk of produced hadrons. For such a back-extrapolation
to be useful, the analysis of the measured hadron spectra should
permit a more or less complete reconstruction of both the geometry and
the dynamics of the source at the point of freeze-out.

For this reason the last few years have seen a growing effort to
understand the final state of heavy ion collisions quantitatively from
the measured hadron spectra. In this project a crucial role is played
by two-particle momentum correlations. Such correlations, whether due
to final state interactions or quantum statistics, are sensitive to
the {\em phase-space distribution} of the particles in the source at
freeze-out and thus carry information not only on the momentum-space
structure of the source, but also on its size and shape in coordinate
space. In particular, they are sensitive to correlations between
the momenta and positions of the particles at freeze-out such as those
which are generated by the collective expansion of the hot and dense
collision zone. This dynamical aspect is very important since, once
identified and quantitatively analyzed, it carries valuable
information about the early equilibration processes and the resulting
build-up of thermodynamical (isotropic) pressure in the reaction zone,
as well as on the time-integrated action of the equation of state of
the quark-gluon and/or hadronic matter during the expansion stage.

The production of deuterons, antideuterons and larger nuclear clusters
via coalescence of (anti-)nucleons at freeze-out can be viewed as a
particular type of phase-space correlation among the particles in the
final state. The similarity of the physics of coalescence and of other
types of final state momentum correlations was stressed before by
Mr\'owczy\'nski \cite{mrow} who pointed out that similar source
information can be extracted from the analysis of both types of
correlations. In the present paper we carry the analogy significantly
farther by extending the investigation to systems with strong
collective dy\-na\-mics. We show in particular that the same ``lengths
of homogeneity in the source'' \cite{MS88} which can be extracted
from two-particle Bose-Einstein correlations (Hanbury Brown -- Twiss
(HBT) interferometry \cite{HBT}) determine the coalescence probability 
in the cluster formation process. This means that the effects of
source expansion enter the ``HBT radii'' and the cluster yields in a
similar way. For this reason nuclear cluster spectra provide an
important and complementary check for the source reconstruction from
single-particle spectra and two-particle momentum correlations of
elementary hadrons. 

As a point in case we demonstrate that we can successfully reproduce
the measured cluster yields and spectra for deuterons, antideuterons,
tritons and $^3$He in Pb+Pb collisions at the CERN SPS
\cite{kabana,kabanapriv,murray,murraypriv} with a source model whose
parameters were determined \cite{urs,NA49PbPb} from a combined
analysis of pion spectra and two-pion correlations. We further show
that the shape (inverse slope) of the measured deuteron transverse
mass spectra \cite{murray,murraypriv}, when compared with that of the
proton spectra, yields additional constraints on the shape of the 
transverse source distribution in space which remove a specific 
ambiguity left open by the analysis of pion spectra and correlations.

Historically, cluster formation has been characterized in terms of the 
``invariant coalescence factor'' $B_A$, defined through the invariant
cluster momentum spectra via the equation 
 \begin{equation}
 \label{f89}
   E_A \frac{dN_A}{d^3 P_A}  =  B_A  
   \left( E_{\rm p}\frac{dN_{\rm p}}{d^3 P_{\rm p}} \right)^Z
   \left( E_{\rm n}\frac{dN_{\rm n}}{d^3 P_{\rm n}}
   \right)^N_{\big|{P_{\rm p}=P_{\rm n}=P_A/A}}  . 
 \end{equation}
It relates the cluster spectrum to the invariant momentum spectra of
the coalescing nucleons at the same velocity. The correct physical
interpretation of $B_A$ has been a long-standing problem
\cite{mrow,butler,schwarzschild,gutbrod,bond,mekjian,kapusta,sato,remler,%
csernai,mrow1,dover,danbertsch,pd,leupold,llope,nagle}.
The relative fragility of the nuclear clusters implies that only
nucleons with small relative momenta contribute to cluster formation;
this led early researchers \cite{butler,schwarzschild,gutbrod} to
an interpretation of $B_A$ in terms of a {\em momentum space
  coalescence volume}, parametrized in terms of a maximal relative
momentum $p_0$ between the coalescing nucleons. This interpretation
appeared to be confirmed by the approximate constancy of the $B_A$
values observed in heavy ion collisions at the BEVALAC with beam
momenta up to about 1~GeV/nucleon, independent of the beam energy and
the size of the colliding nuclei. In the sudden approximation
\cite{bond} and the thermodynamic \cite{mekjian} models, on the other
hand, $B_A\sim V^{A-1}$ is inversely related to the fireball volume in
{\em coordinate space}. In such a picture the decrease of the $B_A$
values with increasing beam energy, observed in nuclear collisions
with large ions at beam energies above 1 GeV/nucleon (for a recent
compilation see \cite{klingenberg}), can be easily understood in terms
of collective expansion of the collision zone before break-up. Good
reviews of these early approaches can be found in
\cite{kapusta,csernai,nagle}.  

Later work began to stress more explicitly the {\it phase-space}
aspects of the coalescence process, starting from a quantum mechanical 
approach based on the density matrix of the source or the equivalent
Wigner function formalism \cite{sato,remler,mrow1,danbertsch,pd} and
its classical phase-space analogues \cite{dover,leupold}. In this
approach the size of the cluster itself enters as an additional
dimensionful quantity into the calculation of $B_A$, and it allows to
address the problem of energy-momentum conservation in the coalescence 
process. 

None of the existing model calculations accounts, however, properly
for {\em dynamical expansion} of the source and the resulting 
correlations between the momenta and positions of the particles at
freeze-out. These correlations can be included numerically by applying
a density matrix or Wigner function based ``coalescence afterburner''
to the output of classical microscopic phase-space simulations like
the Intranuclear Cascade \cite{remler}, ARC \cite{kahana}, or RQMD
\cite{nagle,mattiello,sorge}. While such numerical simulations may
succeed or fail to reproduce the experimental data, in neither 
case they provide a clear conceptual understanding of the physics
entering into the calculation of $B_A$. In this paper we present an
analytical approach, based on an explicit source parametrization,
which provides such an understanding. Like the static model
calculations before, it is based on a quantum mechanical density
matrix (Wigner function) approach. It thus allows for a proper
treatment of energy-momentum conservation and to exhibit the
dependence of $B_A$ on the phase-space structure of the source,
i.e.\ on its longitudinal and transverse size and collective flow, and
on the internal cluster structure.

The paper is organized as follows: In Sect.~\ref{sec2} we review
shortly the classical thermal+flow model approach to cluster
production in which clusters are viewed as pointlike elementary
particles without internal structure which are produced by thermal
emission from an expanding source. While poorly justified
theoretically, this approach has enjoyed considerable phenomenological 
success. The results of Sect.~\ref{sec2} will serve as a benchmark for 
the discussion of the quantum mechanical density matrix approach
presented in the following sections. We will discover strong formal
similarities which simultaneously provide a theoretical justification
for the phenomenological success of the thermal approach and yield
an interpretation of the thermal model parameters for the cluster
spectra in terms of the phase-space characteristics of the underlying 
nucleon source. This is described in Sects.~\ref{sec3} and \ref{sec4}
where we discuss the quantum mechanics of the coalescence process and
calculate a quantum mechanical correction factor for the classical
thermal cluster spectrum. We show that the latter is given by a very
simple expression involving only the size of the cluster and the
``homogeneity radii'' of the source which can also be measured with
HBT interferometry. In Sect.~\ref{sec5} we extend this discussion from
two-nucleon clusters ($d$ and $\bar d$) to three-nucleon clusters, $t$
and $^3$He. In Sect.~\ref{sec6} we express the $B_2$ value in terms of 
of the HBT radii and show how cluster yields can be used for an
analysis of the {\em chemical} composition of the fireball at the
point of {\em thermal} nucleon freeze-out. This complements the
chemical analysis of elementary hadron yields which determines the
(usually much earlier) point of {\em chemical} freeze-out. A
comparison of our results with heavy ion data is presented in
Sect.~\ref{sec7}, and our conclusions are presented in
Sect.~\ref{sec8}.  

We will use natural units $\hbar = c = k_{\rm B} = 1$. We denote by 
$m$, $m_t$, and $p_t$ the rest mass, transverse mass, and
transverse momentum of a nucleon. $M$, $M_t$, and $P_t$
denote the respective variables of a cluster. For a cluster of $A$
nucleons we have in good approximation $M = A\, m$, $P = A\, p$, $P_t
= A\, p_t$, and $M_t = A\, m_t$. We denote by $Y = y - y_{\rm cm}$
the longitudinal rapidity of a particle relative to c.m.\ frame of the
fireball.

\section{Thermal cluster spectra}
\label{sec2}

Clusters are no elementary hadrons. Their binding energies
($B=-2.25$\,MeV for a deuteron) are small compared to typical  
collision energies in the fireball created in a heavy ion collision,
and this makes them very fragile objects. While the spectra of
elementary hadrons decoupling from a thermalized fireball along a
freeze-out hypersurface $\Sigma_{\rm f}(R)$ with normal 4-vector
$d^3\sigma_\mu(R)$ can be described by the Cooper-Frye formula
\cite{cofr}, 
 \begin{equation}
 \label{f10}
   E \frac{d^3 N_i}{d^3 P} =  
   \frac{2J+1}{(2\pi)^3} \int_{\Sigma_{\rm f}} P\cdot d^3\sigma(R)\, 
   f_i(R,P)\,, 
 \end{equation}
where $f_i(R,P)$ is a local equilibrium distribution
 \begin{equation}
 \label{f3}
   f_i(R,P) = \left[
   \exp\bigg(\frac{P\cdot u(R) - \mu_i(R)}{T(R)}\bigg) \pm 1 
 \right]^{-1} 
 \end{equation}
with local temperature $T(R)$, local chemical potential $\mu_i(R)$,
and local flow 4-velocity $u^\mu(R)$ (normalized to $u\cdot u=1$), a
similar expression for nuclear clusters cannot be justified: clusters
do not pre-exist as particles with thermalized momentum distributions
in the collision fireball, but are only created by final state
interactions among the nucleons during the freeze-out process.

The absence of preformed nuclear clusters in the collision zone of a
relativistic heavy ion collision at AGS or SPS energies is easily 
seen: Nuclear fragmentation of the target and projectile nuclei can be
excluded as the origin of cluster production if clusters made from
anti\-nu\-cleons (e.g.\ $\bar d$ or $^3{\bar{\rm H}}$) are selected. For
clusters with positive baryon number near the midrapidity region
it was estimated in \cite{johnson} that the probability for a nuclear
fragment to absorb the necessary longitudinal momentum transfer of
several GeV to several 10 GeV per nucleon without getting destroyed 
can be neglected. On the other hand, by applying Levinson's theorem
\cite{colltheo} to the two-particle terms in the virial expansion of 
the grand canonical potential, it was shown in \cite{dashen,schmidt}
that thermal cluster production inside the fireball is exponentially  
suppressed when the binding energy of the cluster is much smaller than
the fireball temperature.

In a very simple-minded approach to cluster formation at nucleon
freeze-out one may postulate that a cluster of $A=Z+N$ nucleons
with total momentum $P$ is emitted whenever $Z$ protons and $N$
neutrons of identical momenta $P/A$ happen to be at the same place
$R$. This results in a generalized Cooper-Frye formula for clusters:
 \begin{eqnarray}
 \label{f11} 
   E \frac{d^3N_A}{d^3 P} &=&
   \frac{2J_A+1}{(2\pi)^3} \int_{\Sigma_{\rm f}}
   P\cdot d^3\sigma(R)\, 
 \nonumber\\
   &&\qquad \times\ 
   f_{\rm p}^Z(R,P/A)\, f_{\rm n}^N(R,P/A)\, ,
 \end{eqnarray}
where $f_{\rm p,n}$ are given by Eq.~(\ref{f3}). This formula agrees,
up to a quantum mechanical correction factor which will be discussed
below, with the Hagedorn model for cluster production in high energy
$pp$ collisions \cite{hagedorn}. In spite of its naivity it has been
quite successful \cite{hagedorn,sonder}. It will be one of the main
points of the present paper to explain why this is so. For this
purpose we first analyze Eq.~(\ref{f11}) for the specific source model 
which will later form the basis of our quantitative comparison with
data.

\subsection{Analytical model for an expanding fireball}
\label{sec2a}

There is mounting evidence from phenomenological analyses of existing
data \cite{lee,HBT,urs} as well as from microscopic simulations of
the collision dynamics \cite{Bravina} that during the early stages of
a relativistic heavy ion collision the reaction zone approaches a
state of local thermal equilibrium. The resulting thermal pressure
causes a collective expansion of the system via hydrodynamic flow. 
It thus makes sense to parametrize the momentum distributions of the
hadrons just before decoupling by Eq.~(\ref{f3}), with the local flow
velocity 4-vector $u(R)$ describing the average particle velocity at
point $R$ and $T(R)$ parametrizing the random (``thermal'') local
momentum fluctuations around their average value $p^\mu = m u^\mu(R)$. 
The local chemical potential $\mu_i(R)$ and the local fugacity
$\lambda_i(R) = \exp[\mu_i(R)/T(R)]$ parametrize the local density of
particle species $i$ at point $R$.   

For heavy particles (nucleons and nuclear clusters) at chemical
potentials $\mu_i \ll M_i$ (see below) Eq.~(\ref{f3}) can be very well
approximated by a local Boltzmann distribution. For simplicity we will
assume freeze-out at constant temperature, $T_{\rm f}(R) = T =$
const. (General arguments based on the kinetics of the freeze-out
process~\cite{SH94,MH97} show that this is generally a good
approximation.) The fugacity $\lambda_i(R)$ is split into a constant,
particle-specific term $\lambda_i = \exp(\mu_i/T)$, and a common
density profile $\bar H(R)$ for all particle species. This implements
the assumptions of local chemical equilibrium among the various
particle species at freeze-out and simultaneous freeze-out of all
particle species which seem reasonable for nucleons and nuclear
clusters. 

While in Pb+Pb collisions in the initial state neutrons outnumber
protons by a factor of 1.54, the ratio $\rm n/p$ or $\rm
\bar{n}/\bar{p}$ in the fireball at freeze-out is unknown. Due to
particle production an unknown fraction of the initial isospin
asymmetry in the nucleon sector may be transferred to other particle
species. We introduce separate chemical potentials $\mu_{\rm p}$ and
$\mu_{\rm n}$ for protons and neutrons and define the chemical
potential of a cluster by $\mu_A=Z\mu_{\rm p}+N\mu_{\rm n}$. The
values of $\mu_{\rm p}$ and $\mu_{\rm n}$ must be extracted from a fit 
to the data.

We consider only very central (impact parameter $b\approx 0$)
collisions. The fireball is then azimutally symmetric with respect to
the beam axis (``longitudinal'' or $z$ axis), and the transverse
coordinates are conveniently chosen as $\rho=\sqrt{x^2+y^2}$ and the
azimuthal angle $\phi$. Ultrarelativistic kinematics in the beam
direction suggests the longitudinal proper time $\tau = \sqrt{t^2 -
  z^2}$ and the longitudinal space-time-rapidity $\eta=\arth(z/t)$ as
appropriate longitudinal and temporal coordinates:
 \begin{equation}
 \label{f5}
  R^{\mu} = \big(\tau\,\ch\eta,\, \rho\,\cos\phi,\, 
                 \rho\,\sin\phi,\, \tau\,\sh\eta \big) \, .
 \end{equation}
The particle momenta are parametrized by their rapidity $Y =
\arth(P_z/E)$ along the beam direction and their transverse mass  
$M_t=\sqrt{M^2+P_t^2}$:
 \begin{equation}
 \label{f6}
  P^{\mu} = \big( M_t\,\ch Y,\, P_t\, \cos\Phi,\, 
                  P_t\,\sin\Phi, \, M_t\,\sh Y \big)\, .
 \end{equation}
The flow 4-velocity is conveniently parametrized in terms of 
longitudinal and transverse flow rapidities $\eta_l$ and $\eta_t$,
respectively:  
 \begin{equation}
 \label{f7}
 \FL
  u^{\mu}(R) = \big( \ch\eta_l\, \ch\eta_t,\,
                     \sh\eta_t\, \cos\phi,\,
                     \sh\eta_t\, \sin\phi,\,
                     \sh\eta_l\, \ch\eta_t \big)\, ,
 \end{equation}
where ${\rm th}\,\eta_i=v_i$, $i=l,t$, defines the corresponding flow
velocities. In the spirit of Bjorken~\cite{bjork} we assume a scaling
{\em velocity} profile $v_l=z/t$ in the beam direction while taking 
a power-law {\em rapidity} profile in the transverse direction which is
independent of $z$ and $t$:
 \begin{equation}
 \label{f4}
   \eta_l(\tau,\eta,\rho) = \eta, \qquad
   \eta_t(\tau,\eta,\rho) = \eta_f
   \left(\frac{\rho}{\Delta\rho}\right)^\alpha \,. 
\end{equation}
Here $\Delta\rho$ characterizes the transverse size of the fireball
(see below), while $\eta_f$ represents the strength of the transverse
flow; the power $\alpha$ of the transverse flow profile is generally
chosen as $\alpha\!=\!1$, except for some tests with $\alpha\!=\!0.5$
and $\alpha\!=\!2$ as noted in the text. 

As the fireball expands the scattering rate of the particles decreases
until finally the thermalization of the system breaks down and the
particles freeze out. Consistently with the above {\em Ansatz}\/ for 
the expansion flow profile we assume that this happens at a fixed
longitudinal proper time $\tau_0$ and set $\bar H(R) = H(\eta,\rho)
\,\delta(\tau-\tau_0)$. For the longitudinal and transverse shape of
the density profile $H(R)$ we take Gaussians with widths $\Delta\eta$
and $\Delta\rho$, respectively.  

With these ingredients the distribution functions in (\ref{f11}) take
the form 
 \begin{mathletters}
 \label{f8}
 \begin{eqnarray}
 \FL
 \label{f8a}
   &&f_i(R,P) = e^{\mu_i/T}\, e^{-P\cdot u(R)/T} \, H(R),\quad i={\rm
     p,n} 
 \\
 \label{f8b}
   &&H(R) = H(\eta,\rho)
          = \exp\left( - \frac{\rho^2}{2(\Delta\rho)^2} 
                       - \frac{\eta^2}{2(\Delta\eta)^2} 
                 \right) \,.
 \end{eqnarray}
 \end{mathletters}
The density profile is normalized to a total covariant freeze-out
volume $V_{\rm cov}$:
 \begin{equation}
 \label{f9}
   V_{\rm cov} = \int d^4R\, \bar H(R) 
   = (2\pi)^\frac{3}{2}\, (\Delta\rho)^2\, (\Delta\eta)\, \tau_0\,,
 \end{equation}
where $d^4R=\tau d\tau\,\rho d\rho\,d\eta\,d\phi$. For freeze-out at
constant longitudinal proper time, the integration measure in
(\ref{f11}) over the freeze-out hypersurface is given by 
$P\cdot d^3\sigma(R) = \tau_0\, M_t\, \rho\,d\rho\, \ch(\eta-Y)\,
d\eta\, d\phi$.  

\subsection{Non-zero emission duration}
\label{sec2b}

In \cite{urs}, instead of a Cooper-Frye integral over a 3-dim freeze-out
hypersurface, invariant spectra are calculated as a space-time integral
$\int d^4R\, S_i(R,P)$ over an emission function 
 \begin{eqnarray}
 \label{Surs}
   S_i(R,P) &=& \frac{2J_i+1}{(2\pi)^3}\, M_t\, \ch(\eta-Y)\, 
              e^{(\mu_i-P\cdot u(R))/T}
 \nonumber\\
   &&\qquad\qquad\times\, \tilde H(\eta,\rho)\ J(\tau) \,.
 \end{eqnarray}

[$\tilde H(\eta,\rho) = \sqrt{(2/\pi)}\, H(\eta,\rho)$ differs
from (\ref{f8}) only by the normalization. The present choice 
is more convenient for us because it absorbs some constant terms
in the cluster spectra below which would otherwise scale with the
nucleon number. However, it affects the interpretation of the total
fireball volume ($\tilde V_{\rm cov} = \sqrt{2/\pi} V_{\rm cov}$), and
of the fugacity factor $\exp(\mu/T)$. In the case of Ref.~\cite{urs},
$\mu/T$ is the fugacity averaged over the fireball; in the present
case it is the fugacity at $R=(\tau_0,\bbox{R}=\bbox{0}\,)$.] 

The function $J(\tau)$ implements a smearing of the freeze-out hypersurface 
around $\tau_0$; the choice in \cite{urs} is
 \begin{equation}
   J(\tau) = \frac{1}{\Delta\tau\sqrt{2\pi}}\, 
   \exp\left( \frac{(\tau-\tau_0)^2}{2(\Delta\tau)^2}\right) \,.
 \end{equation}
General conditions for $J(\tau)$ are $\int d\tau\, J(\tau)$ $= 1$, $\int
d\tau\, \tau\, J(\tau)$ $= \tau_0$, and $\Delta\tau \ll \tau_0$. The
last one ensures that one can treat $f_i$ and $H$ as $\tau$-independent,
and that freeze-out times $\tau< 0$ play no physical role. 

For single-hadron spectra $J(\tau)$ can be immediately integrated over, 
reducing the space-time integral over the emission function to the
Cooper-Frye form (\ref{f10}) with (\ref{f8}). A non-zero duration of
particle emission ($\Delta\tau>0$) has, however, an effect on
cluster formation and on other two-particle correlations. In
\cite{urs}, two-pion correlation data from Pb+Pb collisions at the SPS
were fitted with $\Delta\tau = 1.5$ fm/$c$, although with considerable
uncertainty~\cite{TH98}. This value appears to be small enough to be
able to neglect the $\tau$-dependence of the parameters in $f_i$ and
$H$; since estimates in~\cite{scheibl} have shown that the effect of
$\Delta\tau>0$ on cluster formation should then be small, we will
continue to use the simpler Cooper-Frye formalism (\ref{f11}) also for
cluster spectra.   

\subsection{Cluster spectra from the model source}
\label{sec2c}

Inserting the expressions from Sec.~\ref{sec2a} into Eq.~(\ref{f11})
one is led to the following integral:
 \begin{eqnarray}
 \label{f12} 
 \FL
   && E{dN_A\over d^3P} = {2J_A+1 \over (2\pi)^3}\, e^{{\mu_{_A}\over T}}\,
   \tau_0 M_t \int d\eta\, \rho\,d\rho\, d\tilde\phi\, \ch(\eta-Y) 
 \nonumber \\ 
   &&\quad\times\,
   e^{-{M_t\ch(\eta-Y)\ch\eta_t(\rho) 
              -P_t\sh\eta_t(\rho)\,\cos\tilde\phi \over T} 
              -{A\, \rho^2 \over 2(\Delta\rho)^2} 
              -{A\, \eta^2 \over 2(\Delta\eta)^2}} .
 \end{eqnarray}
The integration over $\tilde\phi = \phi-\Phi$ can be done
analytically, yielding a modified Bessel function, but this is not
very helpful for further analy\-ti\-cal progress. More useful is the
following observation~\cite{MS88,CSH95b,CL96}: for fireball
temperatures below the deconfinement temperature of about 150\,MeV, the
ratio $M_t/T \gtrsim 6A$ in the exponent is large, and the integral
will receive contributions only from narrow intervals $\eta$ and
$\rho$. We may thus expand the hyperbolic cosine and sine terms in the
exponent, keeping only the leading terms:
 \begin{mathletters}
 \label{f13}
 \begin{eqnarray}
 \label{f13a}
   \ch(\eta-Y)\, \ch\eta_t &\approx& 
      1 + {\textstyle{1\over 2}}(\eta-Y)^2  
        + {\textstyle{1\over 2}}\eta_t^2 \, ; 
 \\
 \label{f13b} 
   \cos\tilde\phi\,\sh\eta_t &\approx& \eta_f {x \over \Delta\rho}\,.  
 \end{eqnarray}
 \end{mathletters}
This decouples the longitudinal and transverse flow, and the
$\eta$-integration can now be done analytically. Note that the
``saddle point approximation'' (\ref{f13}) becomes unreliable for
large $P_t$ and/or $\eta_f$~\cite{WSH96}. The remaining integrations
are easy and give 
 \begin{mathletters}
 \label{f14}
 \begin{eqnarray}
 \label{f14a}
 \FL
   &&E {dN_A\over d^3P} \approx {2J_A+1\over(2\pi)^3} 
     {V_{\rm cov}\,M_t\,e^{\mu_{_A}-M_t\over T} \over
      \left({M_t\over T}\eta_f^2 + A \right) 
      \sqrt{{M_t\over T}(\Delta\eta)^2 +A}} 
 \nonumber\\ 
   &&\quad\times
   \cosh\left({A\,Y\over{M_t\over T}(\Delta\eta)^2 + A}\right)
   \exp\left({(\Delta\eta)^2/2 \over
             {M_t\over T}(\Delta\eta)^2 +A}\right)
 \nonumber\\ 
   &&\quad\times
   \exp\left({P_t^2\,\eta_f^2/(2T^2) \over 
             {M_t\over T}\eta_f^2 +A} 
            -{A\,Y^2\,M_t/(2T) \over {M_t\over T}(\Delta\eta)^2+A}  
       \right)
 \\
 \label{f14b}
   &&\approx {2J_A+1 \over (2\pi)^3}\,e^{{\mu_{_A}-M\over T}}
   \, M_t\, V_{\rm eff}(A,M_t)
 \nonumber\\
   &&\quad\times\,
   \exp\left(-{M_t-M\over T^*}
             -{A\,Y^2\over 2(\Delta\eta)^2}\right) ,  
 \end{eqnarray}
 \end{mathletters}
with
 \begin{equation}
 \label{f15}
   V_{\rm eff}(A,M_t) = 
   {(2\pi)^\frac{3}{2}\,(\Delta\rho)^2\,(\Delta\eta)\,\tau_0
   \over
   \left({M_t\over T}\eta_f^2 + A \right) 
   \sqrt{{M_t\over T}(\Delta\eta)^2 + A}} 
 \end{equation}
and
 \begin{equation}
 \label{f16}
   T^* = T + {M\over A}\, \eta_f^2 \, . 
 \end{equation}
In going from (\ref{f14a}) to (\ref{f14b}) we replaced the terms in
the second line on (\ref{f14a}) by 1 and assumed non-relativistic
transverse cluster velocities, $v_t=P_t/M_t\ll 1$. (Eq.~(\ref{f16})
should therefore {\em not} be applied to pions~\cite{Nix}!). $V_{\rm
  eff}(A,M_t)$ is the effective volume contributing to the emission of
clusters with mass number $A$ and transverse mass $M_t$~\cite{CL96}
(see Sect.~\ref{sec2e}). $T^*$ is the inverse slope (``effective
temperature'') of the transverse momentum spectrum.

\subsection{Slope of the $M_t$-spectrum}
\label{sec2d}

The relation (\ref{f16}) (with $M/A$ replaced by the hadron mass $m$)
was suggested by Nu Xu {\em et al.} \cite{nuxu,bearden} as a basis for
a systematic separation of collective flow ($\eta_f$) from thermal
motion ($T$) using single-particle spectra. It has recently become
very popular, especially since in heavy ion collisions at the AGS and
SPS the measured $m_t$-spectra of most hadronic species seem to follow
it quite nicely by showing, at sufficiently low $p_t$, inverse slope
parameters which rise linearly with the hadron rest mass
\cite{QM97}. This has been interpreted as strong evidence for the
existence of transverse collective flow \cite{lee,nuxu,bearden}. 

When applied to nuclear clusters, however, Eq.~(\ref{f16}) predicts
exactly the same slope for all mass numbers $A$ since $M\sim A$. This
is in contradiction with experiments both at the AGS and the SPS which 
show considerably flatter $m_t$-slopes for deuterons than for protons
\cite{abbott,NA44,na49.roland}. In order to understand the possible
origins of this discrepancy one must remember that Eq.~(\ref{f16})
rests not only on non-relativistic kinematics, but also on the
Gaussian transverse density and linear transverse flow rapidity 
profiles. These were selected in \cite{MS88,HBT,urs,TH98,CL96} for
technical convenience, and we have kept them here for ease of
comparability. It was pointed out, however, by Polleri {\it et al.}
\cite{polleri} that the Gaussian density profile gives too much weight
to the center of the fireball where the transverse flow is weak. Since
according to Eqs.~(\ref{f8a}) and (\ref{f11}) the profile function
$H(\rho)$ enters the expression for nuclear cluster spectra with the
$A^{\rm th}$ power, the region of weak transverse flow receives the more
weight the heavier the cluster; for Gaussian profiles this divides the
effective strength of the transverse flow exactly by a factor $A$
\cite{polleri} and thus causes the $A$-independence of the cluster
slopes. 

As we will show in Sect.~\ref{sec7b} this cannot be compensated by
decreasing the power $\alpha$ in the transverse flow profile (\ref{f4}) 
even though this does lead to a more rapid increase of the transverse
flow rapidity at small values of $\rho$. Phenomenological consistency
with the observed $A$-dependence of the cluster $M_t$-slopes can only
be achieved by selecting density profiles $H(\rho)$ which, when taken
to the $A^{\rm th}$ power, don't lose weight in the relevant region of
large transverse flow. A transverse box profile (or a smooth version
of it) satisfies this requirement and is shown to work well in
Sect.~\ref{sec7b}. 
 
The Gaussian transverse density profile is problematic also for a
different reason: in combination with the transverse flow profile
(\ref{f4}) it leads to acausal behaviour for heavy particles
with large transverse velocities $v_t$. Due to the Boltzmann factor
$\exp(-P\cdot u(R)/T)$ in the integrand of (\ref{f10}) or
(\ref{f11}), which gives strong weight to space-time points
$R$ with $u(R)=P/M$ (the more so the larger $M/T$), such particles 
are emitted mostly from fluid cells at large transverse radii $\rho$,
i.e.\ in the tail of the Gaussian density distribution. Such matter
should not exist, however: a cold Pb nucleus with an rms radius
$\rho_{\rm rms}=5.5\,{\rm fm}$ \cite{pbrms}, corresponding to a
hard sphere radius $\rho_0 = \sqrt{5/3} \,\rho_{\rm rms}$, can expand
in time $\tau_0$ at most to a maximum transverse radius of $\rho_{\rm
  max}=\rho_0+\tau_0$ (even less, if the transverse expansion velocity
is $<c$). Thus, no matter should exist at radii $\rho > \rho_{\rm
  max}$. In our case nucleons with transverse velocities $v_t > 0.6\,
c$ tend to be emitted from causally forbidden regions. (For the
lighter pions the problem is much less severe due to the larger 
thermal smearing.) 

For the Gaussian density profile (\ref{f8b}) we will therefore
restrict our attention to nucleons and clusters with $v_t\le 0.6\,c$,
i.e.\ $1.0\le M_t/M \le 1.25$. For a proper description of clusters
with larger velocities the Gaussian transverse density profile must be
modified, either by cutting it off by hand or by replacing it with a
causally consistent box profile. Unfortunately, this forfeits the
simple analytical expressions (\ref{f14}-\ref{f16}) and the direct
comparability with the published results from HBT analyses of two-pion
correlations \cite{HBT,urs,TH98}. 

As for the longitudinal rapidity spectrum of clusters, we expect 
from Eq.\,(\ref{f14}) for its width a decrease with $1/\sqrt{A}$
compared to nucleons.

\subsection{Effective source volume $V_{\rm eff}$ and relation to HBT}
\label{sec2e}

The Boltzmann factor in (\ref{f8a}) couples the particle momentum to
the flow vector $u(R)$. This causes a correlation between the velocity 
and the spatial coordinates of the particle, with a ``coupling
constant'' $M/T$ which increases with the particle mass. Particles
inside the fireball are thus sorted with respect to their velocities,
and particles of given momentum are localized in regions of the 
fireball where the flow velocity is close to the particle velocity.

Thus only a fraction of the total fireball volume $V_{\rm cov}$ is
able to emit particles with given momentum. It is this ``homogeneity
volume'' $V_{\rm hom}(m_t)$ which is accessible through HBT
measurements \cite{HBT}. The HBT radii ${\cal R}_\parallel(m_t)$ and
${\cal R}_\perp(m_t)$ which can be extracted from the 2-particle
correlation function in the YKP parametrization \cite{YKP} describe the
corresponding longitudinal and transverse {\it lengths of homogeneity} 
in the source. They can be evaluated for the model (\ref{f8}) as
space-time variances of the source using the general expressions  
given in \cite{YKP}. If these variances are evaluated in the
saddle-point approximation (\ref{f13}) one finds \cite{CSH95b,YKP}
 \begin{mathletters}
 \label{f18}
 \begin{eqnarray}
 \label{f18a}
   {\cal R}_\perp(m_t) &=& {\Delta\rho \over 
   \sqrt{1 + {m_t\over T}\eta_f^2}} ,
 \\
 \label{f18b}
   {\cal R}_\parallel(m_t) &=& {\tau_0\,\Delta\eta \over 
   \sqrt{1 + {m_t\over T} (\Delta\eta)^2}} .
 \end{eqnarray}
 \end{mathletters}
These are just the factors occurring in the effective volume
(\ref{f15}) for $A=1$:
 \begin{eqnarray}
 \label{f19} 
   V_{\rm eff}(1,m_t) &=& 
   (2\pi)^{3/2}\, {\cal R}_\parallel(m_t)\, {\cal R}_\perp^2(m_t) 
 \nonumber\\
   &\equiv& (2\pi)^{3/2}\, V_{\rm hom}(m_t) \, .
 \end{eqnarray}
Thus the effective volume $V_{\rm eff}$ in the cluster spectrum
(\ref{f14}) is very closely related to the homogeneity volume
extracted from HBT measurements with pairs of identical hadrons:
 \begin{equation}
 \label{f17}
   V_{\rm eff}(A,M_t) = {V_{\rm eff}(1,m_t) \over A^{3/2}}  
   = \left({2\pi\over A}\right)^{3/2} V_{\rm hom}(m_t)\, .
 \end{equation}
For deuterons this implies an effective volume which is about 1/3 that 
of the nucleons.

Since at AGS and SPS energies $\Delta\eta\gtrsim 1$ (see
Sect.~\ref{sec2f}), the longitudinal flow term $\sim (m_t/T) 
(\Delta\eta)^2$ in ${\cal R}_\parallel$ dominates over the geometric
term $\sim 1$. In the transverse direction the flow term $\sim
(m_t/T)\eta_f^2$ is much smaller, and the two terms compete with each
other, depending on the value of $\eta_f$. Higher temperature
increases, larger transverse flow decreases the lengths of
homogeneity. For fixed $T,\eta_f$ the homogeneity lengths decrease
with increasing transverse mass $m_t$. According to
Eqs.~(\ref{f19},\ref{f17}) this implies for fixed particle momenta a 
decrease of $V_{\rm eff}(m_t)$ with the particle rest mass roughly
like $M^{-3/2}$. Due to the non-negligible geometric contribution in
${\cal R}_\perp$ the decrease is actually somewhat weaker; for realistic
parameters (see below) the combination $M_t V_{\rm eff}$ in
(\ref{f14b}) turns out to be practically independent of $M_t$: $M_t
V_{\rm eff}(A,M_t) \approx M  V_{\rm eff}(A,M)$. 

Eqs.~(\ref{f18}-\ref{f17}) indicate perfect $m_t$-scaling of the HBT
radii and the effective volume. If true, the values of $V_{\rm eff}$
to be used in (\ref{f14b}) for nuclear clusters could be directly
extracted from two-pion HBT measurements at very high $p_t$ such that
the transverse mass $M_t$ would be the same. Unfortunately, this is an 
artifact of the saddle-point approximation (\ref{f13}) \cite{WSH96};
a numerical evaluation shows that $m_t$-scaling of the HBT radii is
broken by transverse flow, albeit weakly \cite{YKP,HBT}. Appropriate
care must thus be taken before using (\ref{f19}) to compare the
effective volume $V_{\rm eff}$ in cluster formation with the
homogeneity volume $V_{\rm hom}$ extracted from HBT measurements with
pions or kaons.

\subsection{Parameters for central P\lowercase{b}+P\lowercase{b}
  collisions} 
\label{sec2f}

We close this section by specifying the model parameters to be used
later in the calculation of cluster yields and spectra. In Pb+Pb
collisions at 158\,GeV per nucleon, mid-rapidity is at $y_{\rm
  cm}=2.91$. For very central  collisions (4-5\% of the total
inelastic cross section $\sigma^{\rm inel}_{\rm tot}$, corresponding 
to impact parameters $b \le 3.5$ fm), the analysis of pion spectra and
two-pion correlations from NA49 \cite{NA49PbPb,schonfelder,appelshauser}
led to the following estimates for the model parameters
\cite{urs,TH98}:  
$\Delta\eta \approx 1.3$, 
$\Delta\rho \approx 7\,{\rm fm}$,
$\tau_0     \approx 9\,{\rm fm}/c$,
$\Delta\tau \approx 1.5\,{\rm fm}/c$,
$\eta_f     \approx 0.35$ and
$T          \approx 130\,{\rm MeV}$.
To estimate the effects of possible errors in $T$ and $\eta_f$, we
will additionally test the ($T,\eta_f$) combinations of 
(100\,MeV, 0.43) and (168\,MeV, 0.28), which also describe the slope of the
single particle $m_t$-spectrum of negatively charged hadrons (mainly
pions), but not the behaviour of the transverse HBT radius
${\cal R}^\pi_\perp(m_t)$. The effects of uncertainties in the
determination of $\tau_0$ and $\Delta\rho$, which may also be
important, have not been studied. From the above parameter values we
calculate a total covariant fireball volume (\ref{f9}) of 
$V_{\rm cov}\approx 9000\,{\rm fm}^3$, homogeneity lengths of 
${\cal R}_\parallel (1\,{\rm GeV}) \approx (3.2 \pm 0.4)\rm\,fm$  and 
${\cal R}_\perp     (1\,{\rm GeV}) \approx (5.1 \pm 0.8)\rm\,fm$, and 
an effective volume for nucleons of 
$V_{\rm eff}(1,m)\approx(0.15\pm0.06)\times V_{\rm cov}$.  

\section{The quantum mechanics of coalescence}
\label{sec3}
\subsection{Reference frames}
\label{sec3a}

So far we have treated nuclear clusters as point-like particles
without internal structure. This is not realistic: the deuteron, for
example, has already an rms radius of almost 2\,fm \cite{weise} which
is not much smaller than the homogeneity radii given above. A proper
inclusion of finite size effects and the internal cluster wave
function in the description of the coalescence process is therefore
mandatory. This requires a quantum mechanical treatment. In this
section we will concentrate on the two-body problem, i.e.\ the
coalescence of (anti)deuterons.

Due to its small binding energy, only nucleons with small relative
momentum can form a deuteron. In the deuteron rest frame (d-frame) the
relative motion of the two nucleons and their coalescence into a bound
state are thus described by non-relativistic quantum mechanics. A
relativistic formulation is needed, on the other hand, to describe the 
motion of the deuterons (i.e.\ of the center of mass of the nucleon
pair) in the fireball rest frame (f-frame).  

In the f-frame, we denote the deuteron's momentum and c.m.\ space-time 
coordinates by $P_{\rm d}=(E_{\rm d},\bbox{P}_{\rm d})$ and $R_{\rm d} =
(R^0_{\rm d},\bbox{R}_{\rm d})$, and the momenta and space-time
coordinates of its two nucleons by $P_\pm$ and $R_\pm = (R^0_\pm,
\bbox{R}_\pm)$. In the d-frame, the four-momenta $q_\pm$ of the two
nucleons are given in terms of their relative momentum $\bbox{q}$ by
\begin{equation}\label{f46}
q_\pm^{\mu} = \left(\sqrt{m^2+\bbox{q}^2},\pm\bbox{q}\,\right)\,.
\end{equation}
Their space-time coordinates $r_\pm=(t_{\rm d}, \bbox{r}_\pm)$ in the
d-frame can be expressed in terms of the d-frame c.m.\ coordinates
$r_{\rm d} = (t_{\rm d}, \bbox{r}_{\rm d})$ 
and their relative distance 
$r^{\nu} = (0, \bbox{r})$ as $r_\pm= r_{\rm d} \pm {1\over 2}r$. 
Note that in the d-frame $r$ has a vanishing zero component,
consistent with the equal-time nature of non-relativistic quantum
mechanics. The deuteron's four-velocity in the f-frame we denote by
$b^{\lambda} = \big(b^0,\, \bbox{b}\, \big) = P_{\rm d}^{\lambda} / m_{\rm d}$.
The matrix $L$ for a boost from the f-frame into the d-frame 
is then given by
\begin{equation}\label{f56}
\FL
  L^{\mu\nu} = \left(\begin{array}{cccc}
    b_0 & b_x & b_y & b_z \\
   -b_x & -1-{b_x^2\over 1+b_0} & -{b_x b_y\over 1+b_0} & 
                                  -{b_x b_z\over 1+b_0} \\
   -b_y & -{b_x b_y\over 1+b_0} & -1-{b_y^2\over 1+b_0} & 
                                  -{b_y b_z\over 1+b_0} \\
   -b_z & -{b_x b_z\over 1+b_0} & -{b_y b_z\over 1+b_0} & 
                                  -1-{b_z^2\over 1+b_0}    
                     \end{array} \right) .
\end{equation}
The matrix $L^{-1}$ for the inverse boost follows by substituting
$b_i\leftrightarrow -b_i$ for $i=x,y,z$. 

We consider the f-frame coordinates $P_\pm$ and $R_\pm$ as functions
of the f-frame c.m.\ coordinates $P_{\rm d}$ and $R_{\rm d}$ 
and the d-frame relative coordinates $q$ and $r$: 
\begin{mathletters}\label{f49}
\begin{eqnarray}\label{f49a}
R_\pm^\mu & = & \left(L^{-1}\right)^{\mu\nu} r_{\pm,\nu}  
                 = R_{\rm d}^\mu \pm {\textstyle{1\over 2}}
                \left(L^{-1}\right)^{\mu\nu} r_\nu\, ,   
\\
\label{f49b}
P_\pm^\mu & = & \left(L^{-1}\right)^{\mu\nu} q_{\pm,\nu} \, .
\end{eqnarray}
\end{mathletters}

For the internal (relative) wave function $\varphi_{\rm d}$ of the
deuteron we will consider both the Hulthen form with the parameters
$\alpha{=}0.23\,{\rm fm}^{-1}$, $\beta{=}1.61\,{\rm fm}^{-1}$,  
and the wave function of a spherical harmonic oscillator with the size
parameter $d{=}3.2\,\rm fm$ ($r\equiv \vert\bbox{r}\vert$): 
 \begin{mathletters}
 \label{f22}
 \begin{eqnarray}
 \label{f22a}
   \varphi_{\rm d}(\bbox{r}) &=& 
   \sqrt{{\alpha\beta(\alpha + \beta) \over 2\pi (\alpha - \beta)^2}}\,
   {e^{-\alpha r} - e^{-\beta r} \over r}\, ,
 \\
 \label{f22b}
   \varphi_{\rm d}(\bbox{r}\,) &=& 
   \left(\pi d^2\right)^{-3/4} \exp\left(- {r^2\over 2d^2} \right) .
 \end{eqnarray}
 \end{mathletters}
While the first choice gives a better description of the deuteron
ground state, the second one is technically advantageous since it
allows for a largely analytic evaluation of the coalescence factor
below. Both wave functions reproduce the measured rms radius of
1.96~fm \cite{weise}. Please note that the variable $\bbox{r}$  
describes the diameter and not the radius of the relative
wave functions such that the rms radius is given by $r_{\rm rms}^2 =
\int d^3r\, \left({\bbox{r}\over 2}\right)^2 \vert\varphi_{\rm d}
(\bbox{r}) \vert^2$.     

\subsection{The density matrix formalism}
\label{sec3b}

According to the rules of statistical quantum mechanics \cite{feynman}
the number of created deuterons with momentum $P_{\rm d}$ is given by
projecting the deuteron density matrix onto the two-nucleon density
matrix in the fireball at the freeze-out time $t_{\rm f}$:   
 \begin{eqnarray}
 \label{f20}
   &&{dN_{\rm d}\over d^3P_{\rm d}} \sim
   {1\over 2!} \int d^3x_1\,d^3x_2\,d^3x_1'\,d^3x_2'\,
   \phi_{\rm d}^*(\bbox{x}_1,\bbox{x}_2)\,
   \phi_{\rm d}(\bbox{x}_1',\bbox{x}_2')\, 
 \nonumber\\
   &&\qquad\times \left\langle \psi^{\dagger}(\bbox{x}_2',t_{\rm f})\,
                       \psi^{\dagger}(\bbox{x}_1',t_{\rm f})\, 
                       \psi(\bbox{x}_1,t_{\rm f})\,
                       \psi(\bbox{x}_2,t_{\rm f})\right\rangle\,.
 \end{eqnarray}
The correct spin and isospin factors will be added later after
introducing Wigner functions. A rigorous ansatz with projection
operators that select the correct spin and isospin state for the
deuteron from the two-particle density matrix can be found in
\cite{mattiello}. 
  
In the detector the deuteron is observed as a free momentum
eigenstate. Its wave function is therefore given by a plane wave for
the c.m.\ motion multiplied by the internal wave function $\varphi_{\rm
  d}$: $\phi_{\rm d}(\bbox{x}_1,\bbox{x}_2) =$ $(2\pi)^{-3/2} 
\exp(i \bbox{P}_{\rm d} \cdot (\bbox{x}_1{+}\bbox{x}_2)/2 )\, 
\varphi_{\rm d}(\bbox{x}_1{-}\bbox{x}_2)$. The two-nucleon density
matrix in the fireball is not known and has to be approximated. We assume  
that at freeze-out the nucleons are uncorrelated:
 \begin{eqnarray}
 \label{f23}
   && \left\langle\psi^{\dagger}(\bbox{x}_2',t)\,
      \psi^{\dagger}(\bbox{x}_1',t)\, 
            \psi(\bbox{x}_1,t)\,\psi(\bbox{x}_2,t)\right\rangle  
 \nonumber\\
   && =
 \left\langle\psi^{\dagger}(\bbox{x}_2',t)\,\psi(\bbox{x}_2,t)\right\rangle\, 
 \left\langle\psi^{\dagger}(\bbox{x}_1',t)\,\psi(\bbox{x}_1,t)\right\rangle\, .
 \end{eqnarray}
The one-particle density matrix can be expressed through the 1-body
Wigner function as \cite{feynman,wigner}
 \begin{eqnarray}
 \label{f25}
   && \left\langle \psi^{\dagger}(\bbox{x}',t)\,\psi(\bbox{x},t)\right\rangle 
 \nonumber\\
   &&\quad = \int \frac{d^3p}{(2\pi)^3}\, f^{\rm W}\left(\bbox{p}\,;t,
        {\textstyle{\bbox{x}{+}\bbox{x}'\over 2}}\right)\,
     \exp\big(i \bbox{p}\cdot(\bbox{x}{-}\bbox{x}')\big).
 \end{eqnarray}
To evalute the integral (\ref{f20}), we introduce new coordinates 
$\bbox{r}_+ = {1\over 2}(\bbox{x}_1+\bbox{x}_1')$,
$\bbox{r}_- = {1\over 2}(\bbox{x}_2+\bbox{x}_2')$, 
$\bbox{\xi} = \bbox{x}_1-\bbox{x}_1'$ $-(\bbox{x}_2-\bbox{x}_2')$ and
$\bbox{\rho}  = {1\over 2}(\bbox{x}_1-\bbox{x}_1'+\bbox{x}_2-\bbox{x}_2')$.
The first two are the classical coordinates of the coalescing nucleons
and will be further reexpressed by $\bbox{r}_{\rm d}$ and $\bbox{r}$.
For the momentum vectors $\bbox{p}_1$ and $\bbox{p}_2$ introduced via
(\ref{f25}) we define the relative momentum $\bbox{q} = {1\over
  2}(\bbox{p}_1 - \bbox{p}_2)$ and the total momentum $\bbox{P}_{\rm
  d} = \bbox{p}_1 + \bbox{p}_2$; the identification with the
deuteron's momentum is due a corresponding $\delta$-function for
3-momentum conservation which arises from the $d^3\rho$ integration.
The $d^3\xi$-integration leads to a term which we recognize as the 
Wigner transform of the internal deuteron wave function,
 \begin{equation}
 \label{f28}
   {\cal D}(\bbox{r},\bbox{q}) = 
   \int d^3\xi\, e^{-i\bbox{q}\cdot\bbox{\xi}}\,
   \varphi_{\rm d}\left(\bbox{r}+{\textstyle{\bbox{\xi}\over 2}}\right)\,
   \varphi_{\rm d}^*\left(\bbox{r}-{\textstyle{\bbox{\xi}\over 2}}\right) ,
 \end{equation}
which has the following normalization property:
 \begin{equation}
 \label{f29}
   \int d^3r \int {d^3q\over (2\pi)^3}\, {\cal D}(\bbox{r},\bbox{q})  
   = \int d^3r\, \varphi^*_{\rm d}(\bbox{r})\, \varphi_{\rm d}(\bbox{r}) 
   = 1 \,.
 \end{equation}
For the deuteron spectrum we thus obtain
 \begin{eqnarray}
 \label{f31}
   {dN_{\rm d}\over d^3P_{\rm d}} &=& {3\over (2\pi)^3} 
   \int d^3r_{\rm d} \int {d^3q\, d^3r \over (2\pi)^3}\,
   {\cal D}(\bbox{r},\bbox{q})
 \nonumber\\
   &&\quad\times 
   f^{\rm W}_{\rm p}\big(q_+,r_+\big)\,
   f^{\rm W}_{\rm n}\big(q_-,r_-\big)\,,
 \end{eqnarray}
where $q_\pm,r_\pm$ are functions of the integration variables
according to Eqs.~(\ref{f49}). 

\subsection{The problem of energy conservation}
\label{sec3c}

Since in Eq.~(\ref{f20}) the two-nucleon density matrix is projected
on the density matrix of an energy eigenstate (the free deuteron),
energy is clearly conserved. On the other hand, if we were to replace
the Wigner functions in (\ref{f31}) by classical distribution
functions of on-shell nucleons (as done e.g.\ in \cite{dover}) we would
violate energy conservation: due to the deuteron binding energy two
free nucleons cannot coalesce into a deuteron without help of a third
body. In other words, one or both of the Wigner functions in
(\ref{f31}) are probed off-shell, and the nucleons in this expression
cannot be considered as free particles. (For free distinguishable
particles in a thermalized system the Wigner function is identical to
the Boltzmann distribution \cite{feynman}.)

A formalism which makes the conservation of energy during coalescence 
more explicit by including the scattering with a third body and which 
thereby gives more control over the replacement of the Wigner
functions by classical distributions was developed by Danielewicz and
Schuck \cite{pd}. Instead of (\ref{f20}) they start from
\cite{remler2}  
 \begin{eqnarray}
 \label{f32}
 \FL
   &&{dN_{\rm d}\over d^3P_{\rm d}} = 
   \lim_{T,T'\to\infty} {1\over T\,T'}\, {1\over 2!}
   \int_{T}^{2T}  \!\!\!dt \int_{T'}^{2T'} \!\!\! dt'
   \int d^3x_1\,d^3x_2\, 
 \nonumber \\ 
   &&\quad \times\, d^3x_1'\,d^3x_2'\, e^{iE_{\rm d} (t-t')} 
   \phi_{\rm d}^*(\bbox{x}_1,\bbox{x}_2)\,
   \phi_{\rm d}(\bbox{x}_1',\bbox{x}_2')\, 
 \nonumber\\
   &&\quad \times 
  \left\langle\psi^{\dagger}(\bbox{x}_2',t')\,\psi^{\dagger}(\bbox{x}_1',t')\, 
         \psi(\bbox{x}_1,t)\,\psi(\bbox{x}_2,t)\right\rangle\, .
 \end{eqnarray}

They then consider the equations of motion for the correlator 
$\left\langle\psi^{\dagger}(\bbox{x}_2',t')\,\psi^{\dagger}(\bbox{x}_1',t')\, 
             \psi(\bbox{x}_1,t)\,\psi(\bbox{x}_2,t)\right\rangle$. 
These include interactions
with third particles which can put one of the two nucleons slightly
off-shell. This nucleon can then form a deuteron with an on-shell
nucleon without violating energy conservation. It is convenient at
this point to introduce a four-vector notation, with momentum
four-vectors which can be either off-shell (indicated by an asterisk)
or on-shell (no asterisk): $p_i =  (E_i,\bbox{p}_i)$, $p_i^* =
(\omega,\bbox{p}_i)$, and $r_i =  (t_i,\bbox{r}_i)$. For the
evaluation of (\ref{f32}) we refer to Ref.~\cite{pd}; the result is
(see Eqs.~(16,19) in \cite{pd})
 \begin{eqnarray}
 \label{f33}
 \FL
   &&{dN_{\rm d}\over d^3P_{\rm d}} = {-3i\over (2\pi)^3} 
   \int d^4r_{\rm d}\, d^3r 
   \int {d^4p_1\over (2\pi)^4}\, {d^3p_2\over (2\pi)^3}\,
 \nonumber\\
   &&\quad \times 
   (2\pi)^4 \delta^4(P_{\rm d}{-}p^*_1{-}p_2)\,
   {\cal D}\left(\bbox{r},
                 {\textstyle{\bbox{p}_1-\bbox{p}_2\over 2}} \right)
 \\ 
   &&\quad \times   
   \Bigl(\Sigma^<_{\rm p}(p_1^*,r_+)\, f^{\rm W}_{\rm n}(p_2,r_-)
            + \Sigma^<_{\rm n}(p_1^*,r_+)\, f^{\rm W}_{\rm p}(p_2,r_-)
   \Bigr),
 \nonumber
 \end{eqnarray}
where the terms $\Sigma^<_{\rm p} f^{\rm W}_{\rm n}$ and 
$\Sigma^<_{\rm n} f^{\rm W}_{\rm p}$ give equal contributions.
The rate $\Sigma_{\rm N}^<(p^*,x)$ for producing a nucleon with off-shell
four-momentum $p^*$ at point $x$ is given by the scattering of the
nucleon with a third particle: 
 \begin{eqnarray}
 \label{f34}
   &&-i \Sigma_{\rm N}^<(p^*,x) = \sum_j \int {d^3q\over (2\pi)^3}\, 
   {d^3p'\over (2\pi)^3}\, {d^3q'\over (2\pi)^3}
 \nonumber\\ 
   &&\qquad\times\
   (2\pi)^4 \delta^4(p^*{+}q{-}p'{-}q')\, 
   |M_{{\rm N}j \rightarrow {\rm N}j}|^2
 \nonumber\\
   &&\qquad\times\
   f^{\rm W}_{\rm N}(p',x)\, f^{\rm W}_j(q',x)\, 
   \big(1 \pm f^{\rm W}_j(q,x)\big)\, .
 \end{eqnarray}
Ref.\,\cite{pd} considers a gas of nucleons only, and therefore there
is no summation over particle species $j$ as in (\ref{f34}). Our
fireball, however, mainly consists of pions, and all particle species
with large cross sections with nucleons must be accounted for in the
production rate.  

With the 4-dimensional $\delta$-functions in (\ref{f33},\ref{f34})
now explicitly accounting also for energy conservation, we proceed to
approximate the Wigner distributions $f_i^{\rm W}$ in these equations
by the classical thermal distribution functions (\ref{f8}). Using the
momentum-conserving $\delta$-function $\delta^4(p^*{+}q{-}p'{-}q')$ we
then have the identity 
 \begin{eqnarray}
 \label{f34a}
   &&{e^{-u{\cdot}p'/T} \over e^{u{\cdot}q'/T}\mp 1}
   \left(1\pm {1\over e^{u{\cdot}q/T}\mp 1} \right) 
 \nonumber\\
   &&\quad =
   {e^{-u{\cdot}p^*/T} \over e^{u{\cdot}q/T}\mp 1}
   \left(1\pm {1\over e^{u{\cdot}q'/T}\mp 1} \right) 
 \end{eqnarray}   
and hence
 \begin{eqnarray}
 \label{f38} 
   &&-i\Sigma_{\rm N}^<(p^*,x) \approx f_{\rm N}(p^*,x)\, 
   \sum_j \int {d^3q\over (2\pi)^3}\, f_j(q,x)\, 
 \nonumber\\
   &&\quad \times
   \bigg[\int {d^3p'\over (2\pi)^3}\, {d^3q'\over (2\pi)^3}\,  
   (2\pi)^4 \delta^4(p^*{+}q{-}p'{-}q')
 \nonumber\\ 
   &&\qquad\qquad\times 
   |M_{{\rm N}j \rightarrow {\rm N}j}|^2\,
   \big(1 \pm f_j(q',x)\big) \bigg]\,.
 \end{eqnarray}
The term in square brackets is the scattering cross section times the
relative velocity between (off-shell) nucleons of momentum $\bbox{p}$
and particles $j$ with momentum $\bbox{q}$, averaged over the thermal
distributions: $[\dots] = \langle \sigma_{{\rm N}j}\, v_{{\rm N}j}
(\bbox{p},\bbox{q})\rangle$. The remaining integral over $d^3q$ and
summation over $j$ then gives the total scattering rate or inverse
scattering time of a nucleon with momentum $\bbox{p}$ at point $x$: 
 \begin{equation}
 \label{f38a} 
   -i\Sigma_{\rm N}^<(p^*,x) = 
   {f_{\rm N}(p^*,x)\over \tau_{\rm sca}^{\rm N}(\bbox{p},x)}\, .
 \end{equation}
The production rate $\Sigma_{\rm N}^<(p^*,R)$ is therefore just given
by the (off-shell!) nucleon distribution function $f_{\rm N}(p^*,R)$
divided by the scattering time of nucleons.

A deuteron has twice the scattering rate of its constituent nucleons.
Since a scattering event is likely to break up the deuteron, and since
we have by assumption free streaming particles after freeze-out at
$t_{\rm f}$, the time integration $dt_{\rm d}$ in (\ref{f33}) should
start at $t_{\rm f}-\tau_{\rm scat}^{\rm N}/2$ and end at $t_{\rm f}$.
Further assuming that we may treat the distribution functions as
constant over this time interval, the inverse scattering time in
(\ref{f38a}) cancels against the time integration, and we are left
with  
 \begin{eqnarray}
 \label{f45}
   {dN_{\rm d}\over d^3P_{\rm d}} &=&   
   {3\over (2\pi)^3} \int d^3r_{\rm d} 
   \int \frac{d^3r\, d^3q}{(2\pi)^3}\, 
   {\cal D}(\bbox{r},\bbox{q})
 \nonumber\\
   &&\qquad \times f_{\rm p}(q_+,r_+)\, f_{\rm n}(q_-^*;r_-)\,.
 \end{eqnarray}
The difference between this expression and (\ref{f31}) is that the
Wigner distributions have been replaced by classical distribution
functions (\ref{f8}) one of which is off-shell ($q_-^*$) so that a
deuteron may be formed without violating energy conservation. The
energy component of $q_-^*$ follows from the condition $q_+ + q_-^* =
(m_{\rm d}, \bbox{0}\,)$ implied by the $\delta$-function in
(\ref{f33}):  
 \begin{eqnarray}
 \label{f47}
   q_+^{\mu} &=& \left(\sqrt{m^2+\bbox{q}^2},\,\bbox{q}\right) ,
 \nonumber\\
   q_-^{*\mu} &=& \left(m_{\rm d}-\sqrt{m^2+\bbox{q}^2},\,
                        -\bbox{q}\right) .
 \end{eqnarray}
Eq.~(\ref{f45}) is also a good basis for studying the quantitative
effects of energy conservation on the coalescence process, for example 
by replacing the off-shell momentum $q^*_-$ by its on-shell limit.

\subsection{Coalescence in the highly relativistic fireball}
\label{sec3d}

We now return to the relativistic motion of the particles in the
fireball frame. In (\ref{f31}) or (\ref{f45}) we need the arguments of 
$f(p,x)$ in the f-frame, so we substitute $r_\pm$ and $q_\pm$ by
$R_\pm$ and $P_\pm$ using (\ref{f49}). After performing the $d^3r$ and
$d^3q$ integrations in the d-frame, we reexpress the $d^3r_{\rm d}$
integration over the deuteron's c.m.\ coordinate at time $t_{\rm f}$
in the d-frame in covariant form: 
$E_{\rm d}\, d^3r_{\rm d} = P_{\rm d} \cdot d^3\sigma(R_{\rm d})$. 
The integral $d^3\sigma(R_{\rm d})$
extends over the freeze-out surface $\Sigma_{\rm f}$ which is
characterized by a relation $R^0_{\rm d,f}(\bbox{R}_{\rm d})$ between
the deuteron freeze-out time and point in the f-frame. In this way we 
obtain
 \begin{eqnarray}
 \label{f52}
   E_{\rm d} {dN_{\rm d}\over d^3P_{\rm d}} &=&
   {3\over (2\pi)^3} \int_{\Sigma_{\rm f}} 
   P_{\rm d} \cdot d^3\sigma(R_{\rm d})
 \nonumber\\
   &\times&
   f_{\rm p}\left(R_{\rm d},{\textstyle{P_{\rm d}\over 2}}\right)\, 
   f_{\rm n}\left(R_{\rm d},{\textstyle{P_{\rm d}\over 2}}\right)\,
   {\cal C}_{\rm d}(R_{\rm d},P_{\rm d}) 
 \end{eqnarray} 
with
 \begin{equation}
 \label{f53}
   {\cal C}_{\rm d}(R_{\rm d},P_{\rm d}) = \int {d^3q\, d^3r\over (2\pi)^3}\, 
   {\cal D}(\bbox{r},\bbox{q})\,{f(R_+,P_+)\, f(R_-,P_-)\over
   f^2\left(R_{\rm d},{\textstyle{P_{\rm d}\over 2}}\right)}\,.
 \end{equation} 
Up to the quantum mechanical correction factor ${\cal C}_{\rm d}$ this is
identical with the classical formula (\ref{f11}). 
${\cal C}_{\rm d}(R_{\rm d},P_{\rm d})$ is an
integral over the internal phase-space coordinates and provides a
measure for the homogeneity of the nucleon phase-space around 
the deuteron c.m.\ coordinates $(R_{\rm d},P_{\rm d}/2)$. 
For a homogeneous nucleon phase-space (static and very large systems) 
the second factor under the integral (\ref{f53}) (which we will call 
{\em ``homogeneity factor''}) approaches unity and (\ref{f53}) reduces 
to the normalization integral (\ref{f29}), yielding ${\cal C}_{\rm d}=1$. 

As we will see, the $q$-dependence of the distribution functions
$f(R_+,P_+)$ and $f(R_-,P_-)$ in (\ref{f53}) is much weaker than that
of the deuteron Wigner density ${\cal D}(\bbox{r},\bbox{q})$ which is
peaked near $\bbox{q}=0$. The distribution functions can thus be
pulled outside the $q$-integral, yielding
 \begin{equation}
 \label{f54}
   {\cal C}_{\rm d}(R_{\rm d},P_{\rm d}) \approx \int d^3r\, 
   {f\left(R_+,{\textstyle{P_{\rm d}\over 2}}\right)\, 
    f\left(R_-,{\textstyle{P_{\rm d}\over 2}}\right)\, \over
    f^2\left(R_{\rm d},{\textstyle{P_{\rm d}\over 2}}\right)}\, 
   |\varphi_{\rm d}(\bbox{r})|^2 .
 \end{equation}
This expression is now very similar to the Hagedorn model for cluster
production in $pp$ collisions \cite{hagedorn}. According to Hagedorn, 
the probability of finding a cluster with certain quantum numbers
(mass $M$, spin, etc.) is equal to the probability to find a really
elementary particle of the same quantum numbers, times the probability
${\cal C}$ that the cluster is contained inside the reaction volume
$\Omega$:   
 \begin{equation}
 \label{f94}
   {\cal C} = \int_\Omega |\varphi|^2\, dV \, .  
 \end{equation}
For $pp$ collisions Hagedorn assumed \cite{hagedorn} that the reaction
volume was given by the pion Compton wavelength, $\Omega = 4 \pi /
(3m_\pi^3)$, resulting in ${\cal C}\approx 0.17$ for the deuteron and
${\cal C}\approx 0.28$ for $^3$He or $^3$H. This is consistent with
the homogeneity radii extracted from two-pion HBT correlations in $pp$
collisions which are also of the order of the pion Compton wavelength
\cite{lorstad}.  

\section{The quantum mechanical correction factor}
\label{sec4}

In this section we evaluate the coalescence factor ${\cal C}_{\rm d}$ for the
source model discussed in Sec.~\ref{sec2} and relate it to the
homogeneity radii ${\cal R}_i$ which can be extracted from HBT
measurements.   

\subsection{Integration over relative momenta}
\label{sec4a}

The explicit dependence of the coordinates $R_\pm$ in (\ref{f53}) on 
$\bbox{r}$ is given by (\ref{f49}): 
 \begin{equation}
 \label{f58}
   R^0_\pm = R^0_{\rm d} \pm {\textstyle{1\over 2}}\, 
                         \bbox{b}\cdot\bbox{r},\ 
   \bbox{R}_\pm = \bbox{R}_{\rm d} \pm {1\over 2}\,
   \left(\bbox{r} + {\bbox{b}\cdot\bbox{r}\over 1+b^0}\,\bbox{b}\right) .
 \end{equation} 
From $R_\pm$ one must calculate 
 \begin{mathletters}
 \label{f58a} 
\begin{eqnarray}
 \label{f58a1}
   \eta_\pm &=& {1\over 2}
   \ln{R^0_\pm+R_{\pm z}\over R^0_\pm-R_{\pm z}}\, ,
 \\
 \label{f58a2}
   \rho_\pm &=& \sqrt{R_{\pm x}^2+R_{\pm y}^2}\, ,
 \\
 \label{f58a3}
   \tau_{\pm} &=& \sqrt{(R^0_\pm)^2 - R_{\pm z}^2}\, ,
 \end{eqnarray}
 \end{mathletters}
which enter the density profiles $H(R_\pm)$ and the flow vectors
$u(R_\pm)$. Since the Lorentz transformation (\ref{f49}) mixes the
components of $\bbox{r}$, the functions $\eta_\pm(\bbox{r})$,
$\rho_\pm(\bbox{r})$, $\tau_\pm(\bbox{r})$, and thus the
integrand of (\ref{f53}) are in general complicated functions of the
integration variables $r_x$, $r_y$ und $r_z$. 

The argument $u(R_\pm)\cdot P_\pm$ of the Boltzmann factor can be
evaluated in any frame. It turns out to be most convenient to
transform $u(R_\pm)$ and $u(R_{\rm d})$ from the f-frame to $u_\pm$ and
$u_{\rm d}$ in the d-frame:  
 \begin{equation}
 \label{f59}
   u_\pm^\mu = L^{\mu\nu} u_\nu(R_\pm), \  
   u_{\rm d}^\mu = L^{\mu\nu} u_\nu(R_{\rm d}).
 \end{equation}
Under the assumption that over the effective integration domain the
flow and chemical potential can be approximated as constants we find
 \begin{eqnarray}
 \label{f60}
   &&{f(R_+,P_+)\, f(R_-,P_-) \over 
      f^2\left(R_{\rm d},{P_{\rm d}\over 2}\right)} =
 \nonumber\\
   &&\quad\
   \exp\left( - {q_+{\cdot}u_+ + q_-{\cdot}u_- 
              - 2m\,u_{\rm d}^0 \over T} \right) 
 \nonumber\\
   &&\quad\times
   \exp\left( - {\rho_+^2 + \rho_-^2 - 2\rho_{\rm d}^2 
                 \over 2(\Delta\rho)^2}
              - {\eta_+^2 + \eta_-^2 - 2\eta_{\rm d}^2 
                 \over 2(\Delta\eta)^2} \right) .
 \end{eqnarray}
Since in the d-frame the relative momentum $\bbox{q}$ is small, the
energy components $q_\pm^0$ can be expanded
non-re\-la\-ti\-vi\-sti\-cal\-ly:  
 \begin{eqnarray}
 \label{f62}
   q_+{\cdot}u_+ + q_-{\cdot}u_- & = & 
   m (u_+^0 + u_-^0) + {\bbox{q}^2\over 2m}\,(u_+^0 \pm u_-^0)
 \nonumber\\
   & & - \bbox{q}\cdot(\bbox{u}_+ - \bbox{u}_-)  
     +{B\over 2} (u_-^0 \mp u_-^0) \,,
 \end{eqnarray}
where $B=m_{\rm d}-2m$. The two signs refer to two different
treatments of energy conservation: the lower sign applies if one of 
the two nucleons (here: $q_-$) is off-shell as prescribed by 
Eq.~(\ref{f47}), the upper sign refers to the case where we simply
neglect energy conservation and take both nucleons on-shell as
in Eq.~(\ref{f46}) (in this case the term $\sim B$ vanishes). With
harmonic oscillator wave functions the Wigner density of the deuteron
is a Gaussian, ${\cal D}(\bbox{r},\bbox{q}) = 8\, \exp(-\bbox{r}^2/d^2
-\bbox{q}^2 d^2 )$, and the integration over $d^3q$ in (\ref{f53})
is straightforward: 
 \begin{eqnarray}
 \label{f64} 
   {\cal C}_{\rm d}(R_{\rm d},P_{\rm d}) & = & 
         \int {d^3r\over (\pi d_{\rm eff}^2)^{{3\over2}}}
   \, \exp\left( -{\rho_+^2 + \rho_-^2 - 2\rho_{\rm d}^2 \over
                   2(\Delta\rho)^2} \right)
 \nonumber\\
   &&\ \times\,
      \exp\left( -{\eta_+^2 + \eta_-^2 - 2\eta_{\rm d}^2 \over
                   2(\Delta\eta)^2} \right)  
 \nonumber\\
   &&\ \times\,
      \exp\left( -{\bbox{r}^2\over d^2} 
                 -{m\over T}(u_+^0 + u_-^0  - 2u_{\rm d}^0) \right)
 \nonumber\\
   &&\ \times\,
      \exp\left( -{B_{\rm eff}\over T}
                 +{(\bbox{u}_+ - \bbox{u}_-)^2\over 
                   4T^2 d_{\rm eff}^2 } \right) .
 \end{eqnarray}
Here the two parameters $d_{\rm eff}^2=d^2+(u_+^0 \pm u_-^0)/(2mT)$ 
and $B_{\rm eff}=B (u_-^0 \mp u_-^0)/2$ depend again on how we deal
with energy conservation. 

\subsection{Integration over relative coordinates}
\label{sec4b}

When trying to perform the integration over $d^3r$ one encounters
subtle causality problems which require some discussion. The
integration in (\ref{f52}) runs over the freeze-out hypersurface
$\tau_{\rm d} = \tau_0$. The hypersurface spanned by the $d^3r$
integration at constant time $t_{\rm d}$ in the d-frame is given in
the f-frame by
 \begin{equation}
 \label{f70}
   R^0 = R^0_{\rm d} 
       + (\bbox{R} - \bbox{R}_{\rm d})\cdot \bbox{b}/b^0 \,,
 \end{equation}
which follows by substituting $\bbox{b}\cdot\bbox{R}_\pm$ into
$R^0_\pm$ in Eq.~(\ref{f58}). This hypersurface is inclined relative to
the freeze-out hypersurface, and it also cuts the $T=|R_z|$
half-planes which are tangent to the light cone originating from the
collision point of the two nuclei. Therefore, when integrating over
$d^3r$ in (\ref{f53}) looking for nucleons which may form a deuteron,
the coordinates $R_\pm$ leave the freeze-out hypersurface and even the 
light cone which contains all produced particles. 

It turns out that this problem is much less severe than it first
appears. First, contributions near the longitudinal light cone are
suppressed by the fact that on the light cone the longitudinal proper
time $\tau=\sqrt{(R^0)^2-R_z^2}$ va\-ni\-shes, leading to a diverging  
flow velocity $u$ whose temporal and longitudinal components are given 
by  $R^0/\tau$ and  $R_z/\tau$, respectively, and, correspondingly, to a 
vanishing Boltzmann factor $\exp(-p{\cdot}u(x)/T)$. This constrains
the effective integration domain in the longitudinal direction to a
region well inside the light cone.

In the transverse directions we must constrain the integration domain
by hand to the inside of the light cone, since (as discussed in 
Sec.~\ref{sec2d}) our source parametrization does not automatically
ensure consistency with causality of the produced particles'\
positions and velocities. Fortunately, this is not a serious
interference: both the deuteron Wigner density and the ``homogeneity 
factor'' in (\ref{f53}) are peaked at small values of $\bbox{r}$. The
latter is the more peaked the more strongly the system expands,
i.e.\ the larger the flow velocity gradients are. As we will see, the
integration domain over $\bbox{r}$ is, to the extent that it is not
already restricted by the finite range of ${\cal D}(\bbox{r},
\bbox{q})$, again given by the homogeneity radii ${\cal R}_i$
discussed in Sec.~\ref{sec2e}. Fortunately, these are smallest for
deuterons with large transverse masses where the causality problems
are expected to be most serious.

In view of their minor numerical effects, we refrain from giving a
detailed account of the technical implementation of these restrictions
into the numerical evaluation of (\ref{f64}), referring instead to
Ref.~\cite{scheibl}. In practice, the following analytical estimates
turn out to be sufficiently accurate even on a quantitative level. 
  
\subsection{Analytic approximation of the correction factor}
\label{sec4c}

Since the measured deuteron momentum spectra do not contain
information on the point of deuteron formation, the relevant 
quantity is the {\em average} correction factor 
 \begin{equation}
 \label{f96}
   \left\langle{\cal C}_{\rm d}\right\rangle(P_{\rm d}) =
   {\int P_{\rm d}{\cdot}d^3\sigma(R_{\rm d})\, 
    f^2\left(R_{\rm d},{P_{\rm d}\over 2}\right)
    \,{\cal C}_{\rm d}(R_{\rm d},P_{\rm d}) \over    
    \int P_{\rm d}{\cdot}d^3\sigma(R_{\rm d})\, 
    f^2\left(R_{\rm d},{P_{\rm d}\over 2}\right)}\,.
 \end{equation}
We will first calculate ${\cal C}_{\rm d}(R_{\rm d},P_{\rm d})$ in analytic
approximation for a special combination of $R_{\rm d}$ and $P_{\rm
  d}$ (given by (\ref{f77}) below) and then argue that the result,
called ${\cal C}_{\rm d}^0$, is actually a very good approximation of
$\left\langle{\cal C}_{\rm d}\right\rangle(P_{\rm d})$. 
Numerical studies \cite{scheibl} confirm the validity of these arguments.

We first concentrate on deuterons with zero transverse momentum which
are at rest in the fluid cell where they are created, i.e.\ whose
four-velocity $b^\mu$ agrees with the flow four-velocity at the
production point, $b=u(R_{\rm d})$:  
 \begin{equation}
 \label{f77}
   b^{\mu} = {1\over \tau_{\rm d}}\big(R^0_{\rm d},\, 0,\, 0,\,
                                       R_{{\rm d}z} \big)\, .
 \end{equation}
In the d-frame the fireball near $R_{\rm d}$ can then be described
non-relativistically as long as the longitudinal flow velocity
gradients are sufficiently small. As long as $d<\tau_{\rm d}$ the
non-relativistic approximation is very good in the relevant region
$r_z\lesssim d$ where ${\cal D}$ is non-zero. Since $b_x{=}b_y{=}0$
the Lorentz transformations (\ref{f49},\ref{f59}) do not mix
longitudinal and transverse directions, and the integrand for the
coalescence factor has the same axial symmetry as the fireball.
Then 
 \begin{mathletters}
 \label{f77a} 
\begin{eqnarray}
 \label{f77a1}
   \eta_\pm &\approx& \eta_{\rm d} \pm {r_z\over 2\tau_{\rm d}}\, ,
 \\
 \label{f77a2}
   \rho_\pm^2 &=& \rho_{\rm d}^2 + {r_x^2+r_y^2\over 4} 
              \pm (r_x R_{{\rm d}x} + r_y R_{{\rm d}y})\, ,
 \\
 \label{f77a3}
   \tau_{\pm} &\approx& \tau_{\rm d} - {r_z^2\over 8\tau_{\rm d}}
 \end{eqnarray}
 \end{mathletters}
after a non-relativistic expansion of (\ref{f58a}). Further, the flow
$u_\pm$ follows from $u(R_\pm)$ by a simple shift in the longitudinal
rapidity: 
 \begin{eqnarray}
 \FL
   &&L^{\mu\nu}\,u_\nu(R) \approx
 \nonumber\\
   &&\ 
   \left( 1{+}{(\eta{-}Y_{\rm d})^2\over 2}{+}{\eta_f^2 
                \rho^2\over 2(\Delta\rho)^2},\,
          {\eta_f R_{x}\over \Delta\rho},\,
          {\eta_f R_{y}\over \Delta\rho},\,
          \eta{-}Y_{\rm d} \right) ,
 \nonumber
 \end{eqnarray}
where we have already used the saddle-point approximation (\ref{f13}). 
For the given values of $d$ and $mT$ the value of $d_{\rm eff}$ in
(\ref{f64}) depends only weakly on $u_\pm^0$. Since 
$u_\pm^0(\bbox{r}){\gtrsim}1$ for $b{=}u(R_{\rm d})$ and small $|\bbox{r}|$, 
we use $d_{\rm eff}^2{\approx}d^2{+}1/mT$ and $B_{\rm eff}{=}0$, or
$d_{\rm eff}^2{\approx}d^2$ and $B_{\rm eff}{\approx}B$, respectively, 
depending on whether or not energy conservation is taken into
account. With these approximations (\ref{f64}) turns into a product of
Gaussian integrals in $r_x$, $r_y$ and $r_z$, with the result
 \begin{mathletters}
 \label{f88}
 \begin{eqnarray}
 \label{f88a}
   {\cal C}_{\rm d}^0 &=& {1\over \gamma_\perp^2\,\gamma_\parallel}\,
             \left({d\over d_{\rm eff}}\right)^3\, 
             \exp\left({B_{\rm eff}\over T}\right)\, ,
 \\
 \label{f88b}
   \gamma_\perp &=& 
   \sqrt{1 + \left({d\over 2\,{\cal R}_\perp(m)}\right)^2 
           - \left({\eta_f\over 2\,T\,\Delta\rho}\right)^2}\, ,
 \\
 \label{f88c}
   \gamma_\parallel &=& 
   \sqrt{1 + \left({d\over 2\,{\cal R}_\parallel(m)}\right)^2 
           - \left({1\over 2\,T\,\tau_{\rm d}}\right)^2}\, .
 \end{eqnarray}
 \end{mathletters}
For the source parameters given in Sec.~\ref{sec2f} the last terms
under the square root in $\gamma_\perp,\,\gamma_\parallel$ are
negligible. They originate from the coupling term $\bbox{q}\cdot 
(\bbox{u}_+{-}\bbox{u}_-)$ in (\ref{f62}) and the resulting term 
$|\bbox{u}_+{-}\bbox{u}_-|^2$ in (\ref{f64}). They would thus be
absent if we had started from the approximation (\ref{f54}) instead of
(\ref{f53}). The smallness of these terms is a good check of the
accuracy of (\ref{f54}). 

The last two factors in (\ref{f88a}) deviate from unity by less than
2\% for temperatures $T$ between 100~MeV and 170~MeV if energy
conservation is properly accounted for; if not, the deviations have
the opposite sign, but remain below 5\%.  

Given the high accuracy of (\ref{f54}), we can use it for a check 
of the sensitivity of ${\cal C}_{\rm d}^0$ on the choice of the internal
deuteron wave function. A numerical integration of (\ref{f54}) 
with the Hulthen wave function (\ref{f22a}) yields values for 
${\cal C}_{\rm d}^0$ which are somewhat larger than those 
for harmonic oscillator wave functions. 
For the source parameters given in Sec.~\ref{sec2f} we obtain for the
harmonic oscillator wave function ${\cal C}_{\rm d}^0=0.81^{+0.03}_{-0.05}$ 
and for the Hulthen form ${\cal C}_{\rm d}^0=0.84^{+0.02}_{-0.04}$ (where the
upper and lower limits indicate the effects from the estimated
uncertainties in ($T,\eta_f$)). 
The numbers in Table~\ref{T1} show that the differences are
sensitive mainly to the transverse and longitudinal flow gradients;
they remain on the level of a few percent for weakly expanding sources 
(leftmost column), become stronger for more rapidly expanding sources 
(rightmost column), and can reach a factor of 2 or 3 in systems with very
small interaction volume ($pp$ collisions). 

The origin of the difference is readily understood: while both
wave functions provide the same rms radius, the maximum of $r^2
|\varphi_{\rm d}(r)|^2$ is at $r\approx \mbox{1.5\,fm}$ for the Hulthen
form and at $r\approx \mbox{3\,fm}$ for the harmonic oscillator. Since
the ``homogeneity factor'' in (\ref{f54}) peaks at small values of
$r$, especially for strongly expanding systems with small homogeneity  
radii, the integral is larger for the more realistic Hulthen
wave function than for the harmonic oscillator one.
 
Numerical calculations show that 
${\cal C}_{\rm d}(R_{\rm d},P_{\rm d})$
varies much less as a function of $R_{\rm d}$ than 
$f^2(R_{\rm d},P_{\rm d}/2)$. 
On the other hand, the particular point $\bar R_{\rm d}$ with 
$u(\bar R_{\rm d}){=}b$ at which ${\cal C}_{\rm d}^0$ was evaluated
corresponds to the maximum of 
${\cal C}_{\rm d}(R_{\rm d},P_{\rm d})$, 
to the maximum of the Boltzmann part of
$f(R_{\rm d},P_{\rm d}/2)$, 
and thus approximately to the maximum of integrand in the numerator 
of (\ref{f96}). We can therefore pull 
${\cal C}_{\rm d}(\bar R_{\rm d},P_{\rm d}) = {\cal C}_{\rm d}^0$ 
in front of that integral and thus have 
$\left\langle{\cal C}_{\rm d}\right\rangle \approx {\cal C}_{\rm d}^0$. 
A numerical check gave 
$\left\langle{\cal C}_{\rm d}\right\rangle(\bbox{P}_{\rm d}{=}\bbox{0}) 
  = 0.79$ 
instead of ${\cal C}_{\rm d}^0=0.81$.

For a boost-invariant source, where every comoving observer has identical 
surroundings, we expect ${\cal C}_{\rm d}(R_{\rm d},P_{\rm d})$ 
to depend only on the difference between the local flow velocity and the 
deuteron's velocity, 
${\cal C}_{\rm d}(R_{\rm d},P_{\rm d})={\cal C}_{\rm d}(u(R_{\rm d})-b)$, 
and $\left\langle{\cal C}_{\rm d}\right\rangle$ to be independent of 
$P_{\rm d}$. 
In our fireball longitudinal boost-invariance is broken by the density 
profile $H(R)$. However, for the Gaussian profile used here 
$\left\langle{\cal C}_{\rm d}\right\rangle$ still turns out to be 
independent of the deuteron's longitudinal rapidity. 
In the transverse direction we have no boost-invariance at all. 
As a consequence we find a slight decrease of 
$\left\langle{\cal C}_{\rm d}\right\rangle$ 
with increasing transverse velocity of the deuteron. 
Deuterons with non-zero transverse velocity see the fireball
Lorentz-contracted in their direction of motion; this decreases the
corresponding length of homogeneity and thus 
$\left\langle{\cal C}_{\rm d}\right\rangle$. 
In the region $m_t/m\le 1.25$, i.e.\ $v_t\le 0.6\,c$, to which we restrict 
our discussion in the case of Gaussian transverse density profiles (see the 
discussion at the end of Sec.~\ref{sec2d}), this effect is small and 
$\left\langle{\cal C}_{\rm d}\right\rangle$ is approximately constant.

We can summarize the results of this section in the following
approximate formula for the quantum mechanical correction factor in
terms of the deuteron size parameter $d$ and the longitudinal and
transverse lengths of homogeneity for nucleons:
\begin{equation}\label{f97}
\left\langle{\cal C}_{\rm d}\right\rangle \approx {1\over
  \left( 1 + \left(\displaystyle 
                   \frac{d}{2\, {\cal R}_\perp(m)}\right)^2 \right)
  \sqrt{ 1 + \left(\displaystyle 
                   \frac{d}{2\, {\cal R}_\parallel(m)}\right)^2}} \,.
\end{equation}
This expression does not depend on the longitudinal rapidity and, for
small transverse velocities, only weakly on the transverse momentum of 
the deuteron. Applying this expression to p+p collisions and inserting
correspondingly for the homogeneity lengths about 1 fm each one obtains 
$\left\langle{\cal C}_{\rm d}\right\rangle \approx 0.15$ 
in good numerical agreement with Hagedorn's value (\ref{f94}). 

\section{Larger clusters: $^3_1$H and $^3_2$He}
\label{sec5}

In this section we give a quick estimate of the quantum mechanical
correction factor for clusters made of three nucleons, i.e.~for $^3$H
and $^3$He. Clearly, for three nucleons the internal wave function and
its Wigner transform is much more complicated than for deuterons. For
the three nucleons with coordinates $\bbox{r}_1$, $\bbox{r}_2$ and
$\bbox{r}_3$, we introduce the c.m.\ coordinates 
$\bbox{R} = (\bbox{r}_1{+}\bbox{r}_2{+}\bbox{r}_3)/3$ and the relative
coordinates $\bbox{\rho} = (\bbox{r}_1{-}\bbox{r}_2)/\sqrt{2}$ and
$\bbox{\lambda} = (\bbox{r}_1{+}\bbox{r}_2{-}2\bbox{r}_3)/\sqrt{6}$. 
With this choice we have $\bbox{r}_1^2{+}\bbox{r}_2^2{+}\bbox{r}_3^2  
  = \bbox{R}^2{+}\bbox{\rho}^2{+}\bbox{\lambda}^2$ and
$d^3r_1\,d^3r_2\,d^3r_3 = 3^\frac{3}{2}\,d^3R\,d^3\rho\,d^3\lambda$.
As before we approximate the internal wave function by a spherical
harmonic oscillator solution \cite{aerts}: 
 \begin{equation}
 \label{f99}
   \varphi(\bbox{r}_1,\bbox{r}_2,\bbox{r}_3) = 
   \left(3 \pi^2 b^4\right)^{-3/4}
   \exp\left(-{\bbox{\rho}^2{+}\bbox{\lambda}^2\over 2 b^2}\right) .
 \end{equation}
This wave function is normalized and has the rms radius $b$:
 \begin{mathletters}
 \label{f99a}
 \begin{eqnarray}
 \label{f99a1}
   &&\int 3^{3/2}\, d^3\rho\, d^3\lambda 
     |\varphi(\bbox{r}_1,\bbox{r}_2,\bbox{r}_3)|^2 =1\, ,
 \nonumber\\   
 \label{f99a2}
   &&r_{\rm rms}^2 = \int 3^{3/2}\, d^3\rho\ d^3\lambda\,
     {\rho^2{+}\lambda^2\over 3}\, 
     |\varphi(\bbox{r}_1,\bbox{r}_2,\bbox{r}_3)|^2 = b^2 \,.
 \nonumber
 \end{eqnarray}
 \end{mathletters}
Both clusters are spin-${1\over 2}$ fermions, and the binding energies 
and rms radii are approximately \mbox{$-8$\,MeV} and \mbox{1.75\,fm}. 
(Note that this rms radius is smaller than for deuterons!) 
$^3$He is somewhat more loosely bound than $^3$H, 
but we neglect this difference here.

To estimate ${\cal C}^0_{\rm t}$ for a $^3$H/$^3$He cluster at rest in
the center of the fireball, with momentum $P=(3m,\bbox{P}{=}\bbox{0})$
and c.m.\ coordinates $R=(R^0,\bbox{R}{=}\bbox{0})$, we evaluated
numerically 
 \begin{eqnarray}
 \label{f100}
   {\cal C}^0_{\rm t} &\approx& \int 3^\frac{3}{2} 
   d^3\rho\, d^3\lambda\,
   |\varphi(\bbox{r}_1,\bbox{r}_2,\bbox{r}_3)|^2\,
 \nonumber\\
   &&\, \times\  
   {f\left(R_1,{P\over 3}\right)\,
    f\left(R_2,{P\over 3}\right)\,
    f\left(R_3,{P\over 3}\right) \over
    f^3\left(R,{P\over 3}\right)}
 \end{eqnarray}
where $R_i\,{=}\,(R^0,\bbox{r}_i)$ are the space-time coordinates of
the three nucleons. The results 
${\cal C}^0_{\rm t}{=}0.78^{ +0.05}_{-0.06}$ for 
$\tau_0{=}\mbox{9\,fm/$c$}$ and 
${\cal C}^0_{\rm t}{=}0.67^{+0.06}_{ -0.07}$ for 
$\tau_0{=}\mbox{6\,fm/$c$}$, respectively, are not 
much smaller than the corresponding values for the deuteron. Whereas
one would generically expect larger flow effects for three than for
two nucleons, the three nucleons in the $^3$H/$^3$He clusters occupy a
smaller and therefore more homogeneous region around their center of
mass. 

As for the deuteron we expect that ${\cal C}^0_{\rm t}$ provides a
good estimate for the average correction factor 
$\left\langle{\cal C}_{\rm t}\right\rangle$. 
A more rigorous calculation must, however, take into account the
binding energy which is larger and thus more important than that of
the deuteron. Unlike deuterons, tritons and $^3$He can be formed 
via an excited state by coalescence of three nucleons without
requiring additional particles for energy-momentum conservation.

\section{Extracting physics from measured cluster spectra}
\label{sec6}
\subsection{The invariant coalescence factor $B_A$}
\label{sec6a}

According to (\ref{f52}), the invariant cluster spectra are given by
(\ref{f14}) multiplied by the quantum mechanical correction factor
$\langle{\cal C}_A\rangle(P)$:
 \begin{eqnarray}
 \label{f93}
   E\,{dN_A\over d^3P} &\approx& M_t\,{2J_A{+}1 \over (2\pi)^3} \,
   e^{{\mu_{_A}-M\over T}}\, \left\langle{\cal C}_A\right\rangle(P)\
   V_{\rm eff}(A,M_t)\,
 \nonumber\\
   &&\times\, 
     \exp\left(-{M_t-M \over T^*} 
               -{A\,Y^2\over 2 (\Delta\eta)^2} \right) .
 \end{eqnarray}
For the invariant coalescence factor $B_A$ defined by (\ref{f89})
we thus find
 \begin{equation}
 \label{f90} 
   B_A  = {2J_A{+}1 \over 2^A} A \left\langle{\cal C}_A\right\rangle 
   {V_{\rm eff}(A,M_t) \over V_{\rm eff}(1,m_t)}   
   \left({(2\pi)^3 \over m_t V_{\rm eff}(1,m_t)}\right)^{A-1}.
 \end{equation}
The factor $A$ arises from $M_t/m_t^A=A/m_t^{A-1}$. (In case of a
static, non-expanding fireball the homogeneity volume $V_{\rm eff}$ in
this expression would be replaced by the total fireball volume $V_{\rm
  cov}$ \cite{mekjian}.) With $V_{\rm eff}$ given by
Eqs.~(\ref{f19},\ref{f17}), we can write $B_2$ as 
 \begin{equation}
 \label{f91}
   B_2 = {3\,\pi^{3/2}\left\langle{\cal C}_{\rm d}\right\rangle \over 2m_t\, 
          {\cal R}_\perp^2(m_t)\,{\cal R}_\parallel(m_t)} \, .
 \end{equation}
Note that the last exponential factor in (\ref{f93}), which depends 
strongly on $M_t$ and $Y$, has cancelled in the ratio. 
With $\left\langle{\cal C}_{\rm d}\right\rangle$ given by (\ref{f97}), 
$B_2$ can thus be expressed completely in terms of the deuteron size $d$ 
and the homogeneity lengths (``HBT radii'') 
${\cal R}_\perp,\,{\cal R}_\parallel$.

Eq.~(\ref{f91}) implies that, in the model of Sec.~\ref{sec2}, $B_A$
is almost momentum independent: both $m_t V_{\rm eff}(1,m_t)$ and
$\left\langle{\cal C}_A\right\rangle$ 
depend only very weakly on $m_t$ and $Y$. 
An important precondition for this weak $m_t$-dependence of $B_2$ is, of
course, the cancellation of the $\exp(-M_t/T^*)$-factors; as discussed
in Sec.~\ref{sec2d}, the latter is due to the Gaussian form of the
transverse density profile in $H(R)$ which according to (\ref{f16})
causes identical inverse slope parameters $T^*$ for all clusters. We
have mentioned before that this is inconsistent with the measurements
(see Sec.~\ref{sec7b} below), and that more box-like transverse
density profiles are phenomenologically preferred. In this case 
Eq.~(\ref{f91}) must be amended as follows:
 \begin{equation}
 \label{f91a}
   B_2 = {3\,\pi^{3/2}\left\langle{\cal C}_{\rm d}\right\rangle \over 2m_t\, 
          {\cal R}_\perp^2(m_t)\,{\cal R}_\parallel(m_t)} \,
   e^{2(m_t-m)\left({1\over T^*_{\rm p}}-{1\over T^*_{\rm d}}\right)}
   \, .
 \end{equation}
Eqs.~(\ref{f97},\ref{f91a}) are the most important theoretical results
of the present paper.

Since neutrons are hard to measure, experiments usually do not provide 
$B_A$, but rather 
 \begin{eqnarray}
 \label{f98}
   B_A^* &=& E_A {dN_A\over d^3P_A} \left/
   \left(E_{\rm p} {dN_{\rm p}\over d^3P_{\rm p}} \right)^{Z+N}
   \right|_{P_{\rm p}=P_A/A}
 \nonumber\\  
   &=& B_A\, \exp\left({N(\mu_{\rm n}-\mu_{\rm p})\over T}\right).
 \end{eqnarray}
Here possibly different chemical potentials for neutrons and protons
are important: If nucleons and antinucleons have the same temperature, 
flow and freeze-out density distribution, as we have assumed, our
model yields identical coalescence factors $B_A$ and $B_{\bar{A}}$ 
for clusters made of matter and antimatter. For $\mu_{\rm n}
\not=\mu_{\rm p}$ the corresponding values $B_A^*$ and $B_{\bar{A}}^*$
will, however, be different, and they will also differ from $B_A$. 
If the initial neutron excess of the cold Pb nuclei were still present
at freeze-out, we would expect for Pb+Pb collisions $B_{\rm d}^*
\approx 1.5\, B_{\rm d} \approx 2.3\, B_{\bar{\rm d}}^*$, i.e.\ quite
large differences. Of course, it is not likely that the large net
isospin remains in the neutron channel until freeze-out; a considerable
fraction is expected to boil off with other produced particles. 
Nevertheless, there may be
a visible effect of $\mu_{\rm n} \not=\mu_{\rm p}$ on $B_2^*$ and
especially on $B_3^*$ for $^3$H (which contains 2 neutrons), as well
as a characteristic difference in the $B_3^*$ values for $^3$H and
$^3$He.
 
\subsection{Cluster fugacities}
\label{sec6b}

From the ratio of particles to antiparticles the fugacities can be
calculated:  
 \begin{equation}
 \label{f102}
   \left. {E_A {dN_A\over d^3P_A} \over
           E_{\bar A} {dN_{\bar A}\over d^3P_{\bar A}}}
   \right\vert_{P_A=P_{\bar A}} =
   \exp\left( \frac{2 \mu_A}{T} \right) = \lambda_A^2 \,.
 \end{equation}
Using the proton fugacity $\lambda_{\rm p} = \exp(\mu_{\rm p}/T)$ from
the ratio p/$\bar{\rm p}$ and the deuteron fugacity $\lambda_{\rm d} =
\exp\big((\mu_{\rm p}{+}\mu_{\rm n})/T\big)$ from the ratio
d/$\bar{\rm d}$ one can in principle extract the neutron fugacity
$\lambda_{\rm n} = \exp(\mu_{\rm n}/T)$. Once the temperature is known
(see below), the separate chemical potentials of neutrons and protons
can thus be determined. In practice, however, the uncertainties in the
p and $\bar{\rm p}$ spectra from insufficiently well known
$\Lambda$ and $\bar\Lambda$ decay contaminations and the large
statistical error bars on the $\rm\bar{d}$ spectrum limit the
usefulness of such an analysis.

\subsection{Freeze-out temperature from cluster ratios}
\label{sec6c}

The thermal freeze-out temperature for pions was determined from a
simultaneous analysis of their spectra and two-particle correlations,
see Sec.~\ref{sec2f}. The same freeze-out parameters appear to also
describe quite well the proton $m_t$-spectra \cite{kaempfer}.
An independent determination of the proton freeze-out temperature
uses the chemical composition rather than the shape of the momentum
distribution at freeze-out. In particular one can try to analyse the 
ratio  
 \begin{equation}
 \label{f103}
   S_{AA'} = \left. 
   \left( E_A \frac{dN_A}{d^3P_A} \right) \bigg/
   \left( E_{A'}\frac{dN_{A'}}{d^3P_{A'}} \right) 
   \right|_{\bbox{P}_A = \bbox{P}_{A'} = \bbox{0} }  
 \end{equation}
of the invariant spectra at zero momentum of two different types of
clusters (including nucleons, $A{=}1$). To investigate such a
possibility is suggested in particular by the data from the NA52
experiment at CERN \cite{NA52} and from the E864 experiment at the AGS 
\cite{pope} which measure cluster yields only very close to $P_t=0$.
With (\ref{f93}) this ratio is given by
 \begin{eqnarray}
 \label{f104}
   S_{AA'} &=& {2J_A+1\over 2J_{A'}+1}\, \lambda_{\rm p}^{Z_A-Z_{A'}}
   \, \lambda_{\rm n}^{N_A-N_{A'}}\, e^{(A'-A){m\over T}}
 \nonumber\\
   &\times& {\left\langle{\cal C}_A\right\rangle  \over 
             \left\langle{\cal C}_{A'}\right\rangle} \,
   {A\,V_{\rm eff}(A,M_A)\over A'\,V_{\rm eff}(A',M_{A'})}  
 \end{eqnarray}
which for known fugacities $\lambda_{\rm n,p}$ is easily solved for
the temperature $T$. [For antiparticles $Z$ and $N$ must be taken
negative in (\ref{f104}).] 

The main error of such a determination of the temperature does not arise 
from the correction factors $\left\langle{\cal C}_A\right\rangle$, which 
do not vary much in our parameter range, but from the sometimes substantial
uncertainties of the experimental value of $S_{AA'}$ and of the
fugacities. Furthermore, the temperature value determined from the
yield ratio at a certain point in momentum space rather than from the
ratio of total yields depends on model assumptions about the shape of
the spectra. For example, we already mentioned that a box
profile for the transverse density distribution gives better results
for the slope systematics of the transverse cluster spectra than the
Gaussian profile from which (\ref{f93}) was derived. (This change does 
not affect the rapidity spectra.) Adjusting the box radius such that
the same rms radius $\Delta\rho$ is reproduced ($\rho_{\rm box} =
2\,\Delta\rho$), we find instead of (\ref{f104}) the same
expression multiplied by a factor $\kappa_A / \kappa_{A'}$ where
 \begin{equation}
 \label{f104a}
   \kappa_A =  \Bigl( 1 - e^{-{ 2 M_A\eta_f^2\over T}} \Bigr)
   \Bigl( 1 + {T\over m\eta_f^2}\Bigr)\, . 
 \end{equation}
(We used the same saddle-point approximation as in the derivation 
of (\ref{f93}), see Sec.~\ref{sec2}. We also kept the strength of the flow
$\eta_f / \Delta\rho$ fixed when switching from the Gaussian to the 
box profile.) For sufficiently large
transverse flow $\eta_f^2 \gg T/m$ this $\kappa_A$ is just a factor 1 
without any consequences.
In the opposite limit, however,
$\lim_{\eta_f\to 0} \kappa_A = 2 A$, i.e.\ the intercept of the invariant 
spectrum at $P_t=0$ is changed by a factor $2 A$. Similarly, changing the
Gaussian longitudinal density distribution into a box doesn't matter
much for systems with strong longitudinal expansion, but yields another
factor $\sqrt{A}$ for static systems. For static systems this just
compensates the factor $A^{-3/2}$ from Eq.~(\ref{f17}), which is
intuitively correct since for static fireballs with constant density
the effective volume $V_{\rm eff}$ must coincide with the total
fireball volume, independent of $A$.

Parametrizing the last factors in $S_{AA'}$ as 
 \begin{equation}
 \label{f105}
   {\kappa_A \over \kappa_{A'}}\, 
   {A\,V_{\rm eff}(A,M_A)\over A'\,V_{\rm eff}(A',M_{A'})} =
   \left({A\over A'}\right)^\chi\, ,
 \end{equation}
we thus find find $-{1\over 2} \lesssim \chi \leq 1$, with the upper
limit reproducing the behaviour for a static, homogeneous fireball and
the lower limit corresponding to  rapidly expanding systems or systems
with a Gaussian transverse density profile. For the source parameters
given in Sec.~\ref{sec2f} we expect to be close to $\chi=-{1\over 2}$. 
However, in order to allow for this kind of model-dependence,
freeze-out temperatures extracted from $S_{AA'}$ should be plotted
against the power $\chi$. 

\section{Comparison with experiment}
\label{sec7}

In this section we discuss the presently available cluster data from
Pb+Pb collisions at the CERN-SPS, taken by the NA52 and NA44
collaborations. The NA52 experiment does not trigger on collision
centrality and measures particles close to $P_t=0$, with different 
rigidities $P/Z$ of the spectrometer magnets providing data at various
rapidities. Recently, the minimum bias data were re-analyzed to
extract the centrality of the collisions. Although the minimum bias
data contain information on p, d, $^3$H, $^3$He, $\bar{\rm p}$,  
$\bar{\rm d}$ and $^3\bar{\rm He}$ \cite{NA52}, their analysis in the
framework of the present model is difficult: the assumed axial
symmetry of the source applies only to central collisions, and also
the model parameters given in Sec.~\ref{sec2f} were extracted from
very central Pb+Pb collisions. For this reason only the impact
parameter selected data from the 4\% most central Pb+Pb collisions
will be really useful for us. At the Quark Matter '97 conference
preliminary $\Lambda,\bar{\Lambda}$-corrected particle ratios $\rm
\bar{p}/p$, $\rm \bar{d}/d$, $B_2\rm (d/p^2)$ and $B_2\rm
(\bar{d}/\bar{p}^2)$ were presented for $y=3.75$ as a function of
centrality \cite{kabana}. They were recently amended by a correction 
to the proton spectra from $\Delta$-decays \cite{kabana2}.

From NA44 we have the $M_t$ spectra of p, d and $B_2\rm (d/p^2)$
\cite{murray,murraypriv} as well as an $M_t$-integrated ratio 
${\rm \bar{d}/d} \approx (0.4-1.5)\cdot 10^{-3}$ \cite{sakaguchi} in
the rapidity range $1.9 \le y \le 2.3$ for the 20\% most central
collisions. 

Finally, we have data for $B_2$ and $B_3$ from Au+Au collisions at the 
AGS with similar centrality (4\%\,$\sigma_{\rm tot}$), but for the
considerably lower beam energy of 11.5\,GeV/nucleon \cite{johnson}. 
Comparison with these data requires a readjustment of the source
parameters. 

A full comparison between theory and data is possible only for the
cluster spectra taken in very central (about 4\%\,$\sigma_{\rm tot}$) 
Pb+Pb collisions, since the model parameter set of Sec.~\ref{sec2f}
describes this class of collisions only. The one data point at this
centrality provided by NA52 can therefore only be used to test the
absolute normalization of (\ref{f93}) or the value (\ref{f90}) of
$B_A$ and does not give information on the spectral behavior. On the
other hand, the $M_t$-shape of (\ref{f93}) can be tested qualitatively
with the NA44 data although, due to the weaker centrality cut, the
normalization and the slope parameters may be somewhat different from
what we expect on the basis of the parameters from Sec.~\ref{sec2f}.

\subsection{NA52 data}
\label{sec7a}

\begin{figure}

\vspace*{\fill}\hspace*{\fill}
\epsfxsize=8.5cm \epsfbox{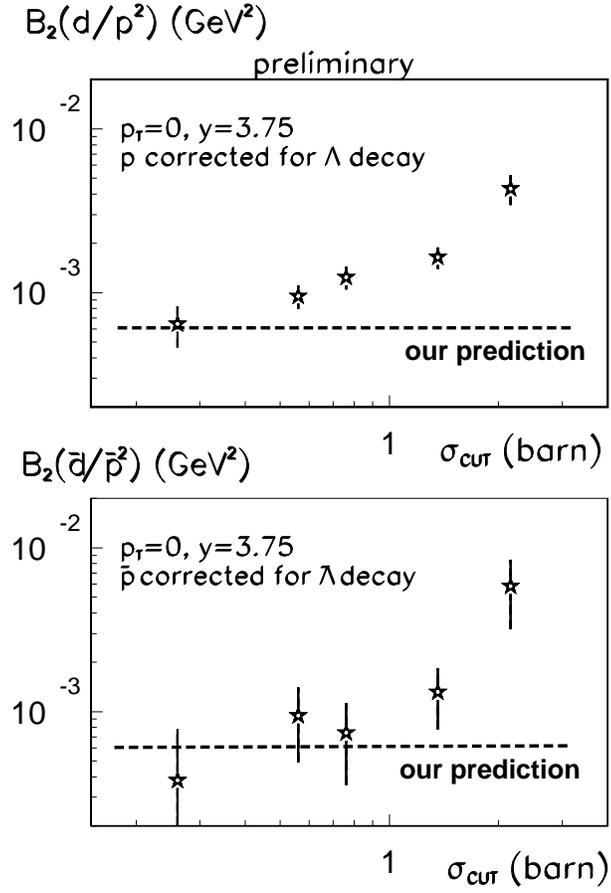} 
\hspace*{\fill}\vspace*{\fill}

\caption{$B_2$ for deuterons and antideu\-te\-rons in 158\,$A$\,GeV
  Pb+Pb collisions as a function of the centrality, measured by
  NA52 \protect\cite{kabana}. (The vertical scale has a systematic
  error of $\sim 40\%$.) With $\sigma_{\rm tot}^{\rm inel} \approx
  8.2\,\rm barn$ in Pb+Pb, the left-most data points correspond to a
  centrality of about 4\%\,$\sigma_{\rm tot}$. The dashed lines show
  our prediction for $B_2$ at 4\%\,$\sigma_{\rm tot}$ (see text).
  \label{na52b2} 
  }

\end{figure}

From the parameters given in section \ref{sec2f} and from our values
$\left\langle{\cal C}_{\rm d}\right\rangle{\approx}0.8$ and 
$\left\langle{\cal C}_{\rm t}\right\rangle{\approx}
 \left\langle{\cal C}_{\rm ^3He}\right\rangle{\approx}0.7$ 
for the quantum mechanical correction
factors, we estimate the invariant coalescence factors $B_A$ in
$4\%\,\sigma_{\rm tot}$ Pb+Pb collisions by using (\ref{f90}):
 \begin{mathletters}
 \label{f92}
 \begin{eqnarray}
 \label{f92a}
   B_2 &\approx& (6^{-2}_{+4})\cdot 10^{-4}\,{\rm GeV^2}\,,
 \\
 \label{f92b}
   B_3 &\approx& (2^{-1}_{+4})\cdot 10^{-7}\,{\rm GeV^4}\, .
 \end{eqnarray}
 \end{mathletters}
Fig.~\ref{na52b2} shows the preliminary NA52 data \cite{kabana} for
$B_2$ as a function of collision centrality, with the most central
collisions on the left. The agreement with our estimate is very good,
both for d and $\rm\bar{d}$, but we should note that it deteriorates 
somewhat if the recently reported $\Delta$-corrected preliminary data 
\cite{kabana2} are used for the comparison. A small neutron excess at 
nucleon freeze-out may explain the sytematically slightly lower $B_2$ 
values for antideuterons (see Eq.~(\ref{f98})). Fig.~\ref{na52b2} shows 
an increase of $B_2$ by up to a factor 8 for more peripheral collisions,
in qualitative agreement with the naive expectation of decreasing
homogeneity lengths and a smaller effective volume in less central
collisions. 

From the preliminary $\rm \bar{p}/p$ and $\rm \bar{d}/d$ ratios
at $y=3.75$ \cite{kabana} and a preliminary d/p ratio \cite{kabanapriv} 
in the most central (4{\%} $\sigma_{\rm tot}$) impact parameter bin 
we tried to determine the freeze-out temperature using (\ref{f104}). 
If we assume a common nucleon fugacity corresponding to $\mu/T=1.5$, 
we obtain $T\approx(144\pm5)$~MeV for Gaussian density profiles 
($\chi{=}-0.5$) and $T\approx(124\pm4)$~MeV for a weakly expanding 
homogeneous fireball ($\chi{=}1$, but still with 
$\left\langle{\cal C}_{\rm d}\right\rangle=0.8$). 
For a larger nucleon chemical potential of $\mu/T{=}1.9$ we 
find $T\approx(135\pm5)\,\rm MeV$ and $T\approx(118\pm4)\,\rm MeV$, 
respectively. The given range of $\mu/T$ at thermal nucleon freeze-out 
and the resulting uncertainties for the temperature are due to the 
experimental errors on the particle ratios. Please note that the 
different values for $\chi$ cause a difference of almost 20\,MeV in 
the extracted temperature. For a purely thermal model with $\chi=1$ and 
$\left\langle{\cal C}_{\rm d}\right\rangle=1$, 
NA52 give a freeze-out temperature of $T\approx(115 \pm 10)\,\rm MeV$ 
\cite{kabana}, taking all uncertainties into account. 
For a transverse box and longitudinal Gaussian density profile 
(as will be motivated below), we have $\chi\approx 0.5$, and using again 
$\left\langle{\cal C}_{\rm d}\right\rangle=0.8$ we obtain
$T\approx(125 \pm 10)\,\rm MeV$. Given the systematic uncertainties, all
these values are consistent with the model parameters given in
Sec.~\ref{sec2f}.  

\subsection{NA44 data}
\label{sec7b}

\begin{figure}

\noindent\parbox{8.0cm}{\epsfxsize=8.0cm \epsfbox{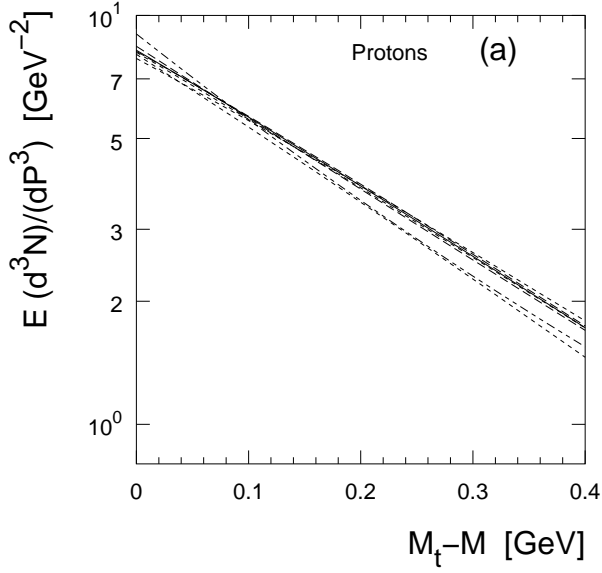} }

\vspace*{0.5cm}
\noindent\parbox{8.0cm}{\epsfxsize=8.0cm \epsfbox{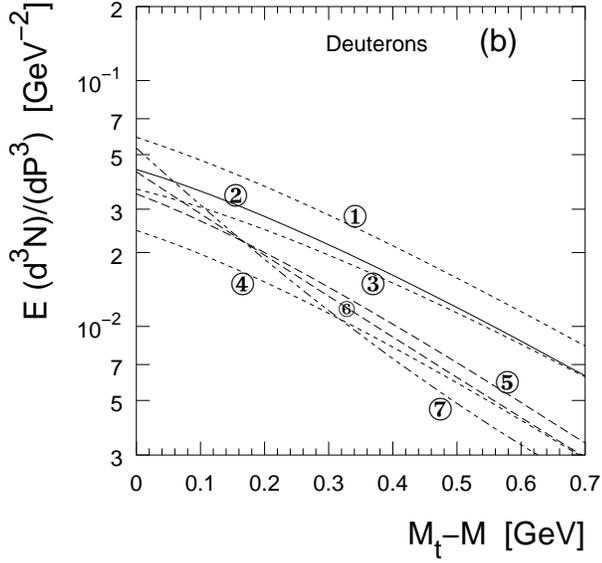} }
\hfill\parbox{8.0cm}{\epsfxsize=8.0cm \epsfbox{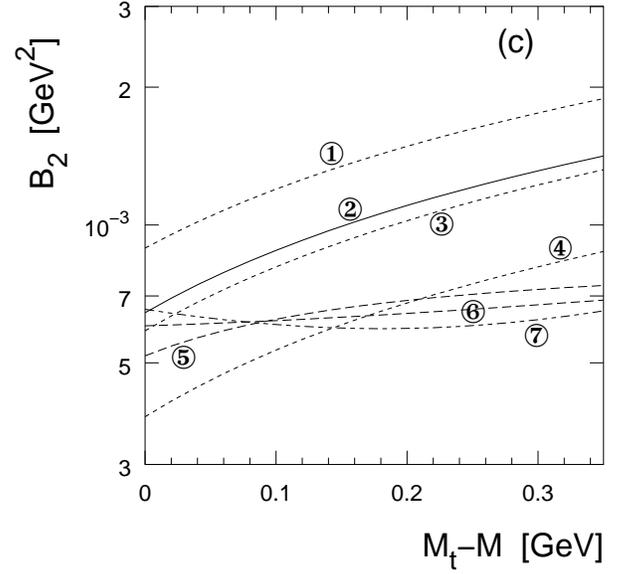}}
\vspace*{\fill}

\caption{Fits to the proton $m_t$-spectrum (a) and predicted deuteron
  $M_t$-spectra (b) and the ratio $B_2=\rm d/p^2$ (c) for different
  models for the transverse density profile. The analysis was
  performed with preliminary data (not shown) from the NA44
  Collaboration \protect\cite{murraypriv} for 20\%\,$\sigma_{\rm tot}$
  Pb+Pb collisions at 158\,$A$\,GeV in the rapidity range $1.9\le y\le
  2.3$. Lines 1--4: box profile for the transverse density with linear
  transverse flow profile; lines 5--7: Gaussian transverse density
  profile with power law transverse flow profiles, with exponents
  $\alpha=0.5$, $\alpha=1.0$ and $\alpha=2.0$. The preliminary data
  lie close to line 2. For details see text. 
  \label{na44fit} 
 }  

\end{figure}

In Fig.~\ref{na44fit} we show the $M_t$ distributions of protons,
deuterons, and $B_2$ from our model. These figures grew out of an
analysis of preliminary data by the NA44 collaboration which were made 
available to us by M.~Murray \cite{murraypriv}, but have so far not
been published (which is the reason why no data are shown). A fit of
the $M_t$-dependence of the invariant proton and deuteron momentum
spectra with the function $\exp(-(M_t-M)/T^*)$ yielded effective
temperatures $T^*_{\rm p}\approx 250$~MeV and $T^*_{\rm d}\approx
350$~MeV, respectively, i.e.\ a considerably larger value for deuterons
than for protons. As a consequence, the measured $B_2$ rises with
$M_t$ by about a factor 2--3 in the transverse mass region
$m_t/m=$1.00--1.25. Higher values of $T^*$ for deuterons than for
protons have been repeatedly observed \cite{abbott,NA44,na49.roland}
and are in clear contradiction with the prediction of identical slope
parameters (\ref{f16}) from a Gaussian source model.

Recently, several possibilities were suggested to account for this
discrepancy \cite{polleri}: a transverse flow profile (\ref{f4}) with
$\alpha=0.5$, a box profile for the transverse density distribution,
and surface emission of clusters. We have analyzed the first two
possibilities.

Starting from (\ref{f12}) with a linear transverse flow profile
$\alpha=1$ and the source parameters of Sec.~\ref{sec2f}, we first
selected different combinations of $\left(T,\, \eta_f,\, 
{\mu\over T}\right)$ which all gave good fits to the proton spectrum
(Fig.~\ref{na44fit}a). The lines for the deuteron and $B_2$ spectra in 
Figs.~\ref{na44fit}b,c follow then self-consistently from the 
parameters used in the proton spectrum without further adjustment.
Line no.\,6 in Figs.~\ref{na44fit}b,c corresponds to the parameter
triplet (140 MeV,\,0.34,\,1.75); it clearly fails to reproduce the
deuteron and $B_2$ spectra. Combinations with larger temperatures and
lower transverse flow rapidities and vice versa lead to deuteron
spectra with the same slope, but somewhat different normalizations. 
The problem of different p and d slopes cannot be resolved in this
way. Please note that there is only a small temperature window for
which the fugacity $\mu/T$ needed to reproduce the normalization of
the proton spectrum compares well with the value $1.5 \lesssim \mu/T
\lesssim 2.0$ from the experimental $\rm \bar{d}/d$ ratio; due to the
high nucleon mass, the nucleon and cluster yields are highly sensitive
to the temperature. The fits tend to under- rather than overestimate
the deuteron yields; this eliminates the simple suggestion that
deuterons from the center of the fireball (which tend to have smaller 
$P_t$) are absorbed by rescattering, leaving only a thin surface shell 
contributing to the observed deuteron yield, because this would
further aggravate the normalization problem.

In a second step, we tested for fixed $T=140\,\rm MeV$ and $\mu/T=1.75$
different transverse flow profiles (\ref{f4}). For $\alpha=0.5$ and
$\alpha=2$ good fits for the proton spectrum can be obtained with
$\eta_f=0.38$ and $\eta_f=0.24$, respectively. The effects of the
modified flow profile on the deuteron and $B_2$ spectra are small:
lines no.~5 correspond to $\alpha=0.5$, which results in a weak, but 
insufficient rise of $B_2$ with $M_t$; $\alpha=2$ (lines no.\,7) made
things worse. 

Finally, we fixed $\mu/T\!=\!1.75$ and
tested transverse box profiles with different radii
$\rho_{\rm box}$ for the function $H(R)$ at different lifetimes $\tau_0$. 
As before, the remaining parameters $T$ and $\eta_f$ are
adjusted to provide good fits to the proton spectrum. Lines no.\,1 to
no.\,4 correspond to the following combinations of
$(\tau_0,\,\rho_{\rm box},\, T,\, \eta_f,\, \mu/T)$,
respectively (values of $\eta_f$ refer to (\ref{f4}) with fixed 
$\Delta\rho=7\,\rm fm$!):\newline 
1. (7~fm/$c$, 10~fm, 142~MeV, 0.40, 1.75); \newline
2. (7~fm/$c$, 12~fm, 136~MeV, 0.34, 1.75); \newline
3. (8~fm/$c$, 12~fm, 134~MeV, 0.36, 1.75); and \newline
4. (9~fm/$c$, 14~fm, 126~MeV, 0.28, 1.75). \newline
All four data sets reproduce the slope of the preliminary deuteron 
and $B_2$ data very nicely. They differ somewhat in the normalization
of the deuteron spectrum. The preliminary data are located between
curves 1 and 3, close to curve 2. 
The parameter choice no.\,4 seems to be the most appropriate with respect
to the Gaussian parameter set of Sec.~\ref{sec2f}, since
$\rho_{\rm box}\!=\! 2\, (\Delta\rho)\! \approx\! \mbox{14\,fm}$ 
for the box radius leads to the same transverse rms radius of the source 
(which is the observable determined from pion interferometry) 
as the Gaussian profile. 
However, it underestimates the yield of the deuterons and the value of 
$B_2$, indicating too large an effective source volume for nucleons (see
Sec.~\ref{sec2e} and Eq.~\ref{f90}). 
Better results are obtained with somewhat smaller values of 
$\rho_{\rm box}$ and $\tau_0$ (curves 1-3 in Figs.~\ref{na44fit}b,c).
This is not necessarily in contradiction to the parameter set of 
Sec.~\ref{sec2f}, which was extracted from very central 
Pb+Pb collisions. For the less central NA44 collisions somewat smaller 
values of $\tau_0$ and/or $\Delta\rho$ may indeed be expected.

Given the preliminary nature of the data we did not make a big effort 
to optimize our fit to the data, e.g. make a combined fit to both proton 
and deuteron spectra. However, as minimum consistency requirements 
the lifetime $\tau_0$ and the flow velocity $\eta_f$ must be large 
enough to allow for the transverse expansion from the cold Pb hard sphere 
radius ($\approx\,7\,\rm fm$) to the freeze-out radius $\rho_{\rm box}$. 
This is an additional reason to reject parameter choice no.\,4.
  
The outcome of this study is that Gaussian transverse density profiles
cannot consistently reproduce both the proton and deuteron spectra,
irrespective of the form of the transverse flow profile. A 
transverse box profile for the density works quite well, both with
linear ($\alpha=1$) and nonlinear ($\alpha=0.5$ and 2) transverse flow
profiles. The source parameters needed to reproduce the deuteron
yields and spectra are consistent with those extracted from HBT
interferometry with pions.

The physical interpretation of these results is that the data require 
more nucleons at higher transverse flow than the Gaussian density
profile can provide even with steep velocity profiles. Other
transverse density profiles like a Woods-Saxon or a doughnut 
profile may work similarly well or even better; on the other hand, the 
HBT ana\-ly\-sis of pion correlations has so far provided no hint for
opaqueness of the source \cite{urs}. A good discussion of different
profiles, based on the analysis of E802 data from Si+Au collisions,
can be found in \cite{polleri}. 

\subsection{Au+Au data from the AGS}
\label{sec7c}

Although the size of a cold Au nucleus is about the same as that of
a cold Pb nucleus, the lower beam energy at the AGS will require
different fireball parameters than those given in Sec.~\ref{sec2f},
even if only collisions taken at the same 4\%~$\sigma_{\rm tot}$
centrality are considered. A few simple checks with our theoretical
results can nevertheless be performed. For E877, Johnson provides for
the invariant coalescence factors in the lowest $P_t$ bin the values  
$B_{\rm d} = (1.5\pm0.5)\cdot 10^{-3}\,{\rm GeV^2}$,
$B_{^3{\rm H}} = (1.19\pm0.29)\cdot 10^{-6}\,{\rm GeV^4}$ and
$B_{^3{\rm He}} = (1.25\pm 0.23)\cdot 10^{-6}\,{\rm GeV^4}$
\cite{johnson}. $B_{\rm d}$ is thus about a factor 2 larger than
expected for central Pb+Pb collisions at the SPS and seen by NA52.
Using (\ref{f90}) and assuming that the correction factors
$\left\langle{\cal C}_A\right\rangle$ do not differ much 
between Pb+Pb and Au+Au collisions, we expect 
 \begin{equation}
  \left(\frac{B_2(\mbox{Au+Au})}{B_2(\mbox{Pb+Pb})}\right)^2 \approx
  \left(\frac{V_{\rm eff}(\mbox{Pb+Pb})}{V_{\rm
        eff}(\mbox{Au+Au})}\right)^2 
  \approx \frac{B_3(\mbox{Au+Au})}{B_3(\mbox{Pb+Pb})} \,.
 \end{equation}
These relations work out quite nicely when the Pb+Pb estimates 
(\ref{f92}) and the above experimental values for Au+Au are
inserted. The E877 value of $B_2$ is consistent with the preliminary
results $B_2= (1.8\pm1.0)\cdot 10^{-3}\,{\rm GeV^2}$ from E878 for
central Au+Au collisions at $y=1.7$ and $B_2= (0.6\pm0.4)\cdot
10^{-3}\,{\rm GeV^2}$ from E864 for 10\%~$\sigma_{\rm tot}$ Au+Pb
collisions at $y=1.9$ \cite{pope}. 

\section{Summary}
\label{sec8}

We have presented an analytic treatment, supported by numerical
checks, of the coalescence of two- and three-(anti)nucleon clusters
in a relativistically expanding fireball. We showed that coalescence
of nuclear clusters and two-particle correlations between the final
state momenta of pairs of identical particles can be characterized by
the {\em same} set of effective source parameters, the ``lengths of
homogeneity''. The only additional scale entering the coalescence
probability is the intrinsic size of the cluster. We made the
connection between coalescence and HBT interferometry explicit with
the help of a simple, but phenomenologically successful
parametrization of the expanding source which was already extensively 
studied in connection with two-pion HBT interferometry. For this model 
most calculations can be done in good approximation analytically.

Starting from the quantum mechanical definition of the deuteron
momentum spectrum as a projection of the two-nucleon Wigner density of
the fireball at nucleon freeze-out on the deuteron Wigner function, 
we used the formalism developed by Danielewicz and Schuck \cite{pd} 
to bring the deuteron momentum spectrum into the form (\ref{f52}) of a
modified Cooper-Frye formula. Except for the ``quantum mechanical
correction factor'' ${\cal C}(R_{\rm d},P_{\rm d})$ this formula looks exactly
like a thermal spectrum of elementary particles with the mass of the
deuteron, where the deuteron phase-space distribution is given by a
product of thermal neutron and proton phase-space distributions.
The factor ${\cal C}_{\rm d}$ is given by the integral (\ref{f53}) 
of the deuteron Wigner density over the homogeneity volume 
of the nucleon source;
for each deuteron momentum $P_{\rm d}$ it is peaked at the value 
$R_{\rm d}$ at which the flow velocity equals the deuteron velocity. 
We showed that the peak value for deuterons at rest in the fireball
frame is a good approximation for $\left\langle{\cal C}_{\rm d}\right\rangle$, 
the average of ${\cal C}_{\rm d}$ over the freeze-out hypersurface, which 
is the relevant quantity for the shape of the deuteron momentum spectrum. 
$\left\langle{\cal C}_{\rm d}\right\rangle$ can be expressed by the very 
simple formula (\ref{f97}) in terms of the ratios between the deuteron size 
$d$ and the longitudinal and transverse lengths of homogeneity for nucleons 
at rest, ${\cal R}_\parallel(m)$ and ${\cal R}_\perp(m)$, respectively. 
In our expanding fireball model $\left\langle{\cal C}\right\rangle$ 
turns out to be essentially momentum-independent for clusters of all sizes, 
and is of order $\left\langle{\cal C}\right\rangle\sim 0.7-0.8$ 
for 2- and 3-nucleon clusters formed in 160\,$A$\,GeV Pb+Pb collisions.

The shape of the deuteron momentum spectrum is thus given by the thermal 
model ansatz (\ref{f11}), with $\left\langle{\cal C}\right\rangle$ leading 
only to a modification of the normalization. 
For our Gaussian source parametrization the single particle spectrum 
can be written, up to trivial factors, as a product (\ref{f14b}) of an 
exponential in $M_t$ with inverse slope (``effective temparature'') $T^*$ 
(\ref{f16}), a Gaussian in rapidity with width $(\Delta\eta)^2/A$, and 
an effective volume $V_{\rm eff}(A,M_t)$ which is again given by the 
homogeneity lengths \cite{CL96}, see (\ref{f17}). 
This allows to express the invariant coalescence factor $B_2$ fully in 
terms of the deuteron size and the lengths of homogeneity, see (\ref{f91}).

Unfortunately, the Gaussian model (Gaussian transverse density profile 
with linear transverse flow rapidity profile) leads to identical
$M_t$-slopes for clusters of all sizes and therefore to $B_A$ values
which are essentially momentum-independent. This is contradicted by
experiment. We found, in agreement with Polleri {\em et al.}
\cite{polleri}, that the flatter deuteron than proton spectra
require a density profile which gives for larger clusters more weight
to regions of larger transverse flow. We found that a transverse box
profile works very well and successfully reproduces the measured
slope of preliminary data \cite{murray,murraypriv} for the deuteron
$M_t$-spectrum in Pb+Pb collisions. The consequence is a
momentum-dependence of $B_2$ which now rises as a function of 
$M_t$ according to the simple generalized expression (\ref{f91a}).

It is interesting to observe that in this way the cluster spectra
provide additional information about the source which cannot be
extracted from HBT measurements. The latter constrain only the rms
radii of the effective source, but not the shape of its spatial
distribution. In the meantime, however, it has been found that a
combined analysis of pion spectra and HBT correlations also prefers a
transverse box profile over a Gaussian one because the former gives a
larger total pion yield than the latter, as required by the data
\cite{TH99}.

\acknowledgements

The authors want to thank S. Kabana, R. Klingenberg, and M. Murray
for discussing with us their preliminary data prior to publication.
This work was supported by GSI, BMBF and DFG.



\vspace*{3cm}

%
%

\begin{table}
\caption{\label{T1} The quantum mechanical correction factor 
          ${\cal C}_{\rm d}^0$
  for Hulthen and harmonic oscillator wave functions calculated
  with Eq.~(\ref{f54}), for different
  fireball parameters at nucleon freeze-out (for details see text).}
\begin{tabular}{l|ccc|ccc}
$\tau_0$ [fm/$c$] & \multicolumn{3}{c|}{9.0} 
                  & \multicolumn{3}{c}{6.0}\\ 
\tableline
$T$ [MeV]  & 168  & 130  & 100  & 168  & 130  & 100\ \ \\ 
$\eta_f$   & 0.28 & 0.35 & 0.43 & 0.28 & 0.35 & 0.43\ \ \\ 
\tableline
Hulthen   & 0.86 & 0.84 & 0.80 & 0.80  & 0.78 & 0.74\ \ \\
harm. osc.& 0.84 & 0.81 & 0.76 & 0.76  & 0.72 & 0.66\ \ \\ 
\end{tabular} 
\end{table}

%
%
\newpage \onecolumn

\newpage \onecolumn

\end{document}